\documentclass[aps,prd,groupedaddress,showpacs,amsmath,amssymb,amsfonts,nofootinbib,longbibliography,showpacs,reprint]{revtex4-1}

\pdfoutput=1

\usepackage{graphicx}
\usepackage{color}
\usepackage{slashed}
\usepackage{tabularx}
\usepackage{hyperref}
\usepackage{ifthen}
\usepackage{dsfont}
\usepackage{braket}
\usepackage{mathtools}
\usepackage{tikz}
\usepackage{microtype}

\def\simge{\mathrel{
     \rlap{\raise 0.511ex \hbox{$>$}}{\lower 0.511ex \hbox{$\sim$}}}}
\def\simle{\mathrel{
     \rlap{\raise 0.511ex \hbox{$<$}}{\lower 0.511ex \hbox{$\sim$}}}}
\def\be{\begin{equation}}
\def\ee{\end{equation}}
\def\bea{\begin{eqnarray}}
\def\eea{\end{eqnarray}}

\newcommand{\tr}{\mathrm{tr}\,}

\newcommand{\ReTr}{\mathrm{ReTr}\,}
\renewcommand{\Re}{\mathrm{Re}\,}
\renewcommand{\Im}{\mathrm{Im}\,}

\renewcommand{\i}{\mathrm{i}}
\renewcommand{\d}{\mathrm{d}}
\newcommand{\e}{\mathrm{e}}

\newcolumntype{L}[1]{>{\raggedright\arraybackslash}p{#1}} %linksbündig mit Breitenangabe
\newcolumntype{C}[1]{>{\centering\arraybackslash}p{#1}} %zentriert mit Breitenangabe
\newcolumntype{R}[1]{>{\raggedleft\arraybackslash}p{#1}} %rechtsbündig mit Breitenangabe

\graphicspath{{.}{figures/}{figures}}

\definecolor{blue}{rgb}{0,0,1}

\definecolor{green}{rgb}{0,1,0}

\definecolor{red}{rgb}{1,0,0}

\begin{document}

\title{Quarks and Triality in a Finite Volume}

\newcommand{\JLU}{Institut f\"ur Theoretische Physik, Justus-Liebig-Universit\"at, 35392 Giessen, Germany}
\newcommand{\HFHF}{Helmholtz Research Academy Hesse for FAIR (HFHF), Campus Giessen, 35392 Giessen, Germany}

\author{Milad Ghanbarpour}\affiliation{\JLU}\affiliation{\HFHF}
\author{Lorenz von Smekal}\affiliation{\JLU}\affiliation{\HFHF}

\begin{abstract}
In order to understand the puzzle of the free energy of an individual quark in QCD, we explicitly construct ensembles with quark numbers  $N_V\neq 0\!\mod 3$, corresponding to non-zero triality in a finite subvolume $V$ on the lattice. We first illustrate the basic idea in an effective Polyakov-loop theory for the heavy-dense limit of QCD, and then extend the construction to full Lattice QCD, where the electric center flux through the surface of $V$ has to be fixed at all times to account for Gauss's law. This requires introducing discrete Fourier transforms over closed center-vortex sheets around the spatial volume $V$ between all subsequent time slices, and generalizes the construction of 't Hooft's electric fluxes in the purge gauge theory. We derive this same result from a dualization of the Wilson fermion action, and from the transfer matrix formulation with a local $\mathbb Z_3$-Gauss law to restrict the dynamics to sectors with the required center charge in $V$.

\end{abstract}

\maketitle

\section{Introduction}
\label{sec::intro}

In this paper we revisit the question of whether and how one can define, in a gauge invariant way, the free energy of a single quark in a finite volume $V$. We expect it to be finite at finite temperature $T$, corresponding to non-vanishing canonical partition functions  with quark number $N_q$, 
\be
Z_c(T,V,N_q) \stackrel{!}{\not=} 0\, ,\;\; \mbox{for} \;\; N_q \not= 0 \!\!\mod 3\,. \label{eq::Zquark}
\ee
These correspond to the so-called non-zero triality sectors, however, and this is therefore in conflict with the naive fugacity expansion relating the canonical to the grand-canonical partition function $Z(T,V,\mu)$. To see this explicitly, one analytically continues the latter to imaginary chemical potential $\mu = \i \theta T$ so that the fugacity expansion becomes a Fourier series,
\be
Z^I(\theta) \equiv Z(T,V,\i \theta T) = \sum_{N_q} \, \e^{\i N_q \theta } \, Z_c(T,V,N_q) \, , \label{eq::ZIFS}
\ee
where the partition function at imaginary chemical potential is of course $2\pi $-periodic in $\theta $ because of the conserved integer particle number $N_q$. In QCD, on the other hand, the Roberge-Weiss symmetry tells us that the periodicity in $\theta $ is in fact only $2\pi/3$ \cite{RobergeWeiss1986,KratochvilaForcrand2006}. This implies $Z^I(\theta + 2\pi/3) = Z^I(\theta) $ and therefore
\begin{equation*}
      Z_c(T,V,N_q) \stackrel{?}{=} 0\, ,\;\; \mbox{for} \;\; N_q \not= 0 \!\!\mod 3\, , 
\end{equation*}
in contradiction with (\ref{eq::Zquark}). The reason for this apparent puzzle is of course that we have tacitly assumed the same boundary conditions for all ensembles, usually periodic up to gauge transformations, in the fugacity/Fourier expansion (\ref{eq::ZIFS}). To properly define the canonical ensembles for quark numbers $N_q \not= 0 \!\mod 3$, we have to account for the center flux of the total fundamental charge inside the finite volume $V$ via Gauss's law. 

In the pure $\mathrm{SU}(3)$-gauge theory with only static fundamental sources, i.e.~infinitely heavy quarks, the corresponding superselection sectors are given by 't~Hooft's electric fluxes. In $d$ spatial dimensions, the partition functions $Z_e(\vec e)$ of ensembles with center flux $\vec e \in \mathbb Z_3^d$ are obtained as $\mathbb Z_3^d$ Fourier transfroms of ensembles $Z_k(\vec k)$ with temporally twisted boundary conditions, labeled by their spatial direction $\vec k$, in the $d+1$ dimensional gauge theory,
\be
Z_e(\vec e) = \frac{1}{3^d} \sum_{\vec k \in \mathbb Z_3^d} \e^{\frac{2 \pi \i}{3} \vec e\cdot\vec k} \, Z_k(\vec k)\,.\label{eq::elf}
\ee
Twist $\vec k = k_x \vec e_x$ in the $x$-direction, for example, with $k_x \in \mathbb Z_3 = \{0,1,2\}$  denotes non-periodic boundary conditions such that the net number of center vortices modulo three through all $(t,x)$-planes of the $d+1$ dimensional torus is given by $k_x$ and fixed. In general, $\vec k \in \mathbb Z_3^d$ twisted boundary conditions imply that the Polyakov-loop $L(\vec x)$, including a non-trivial transition function along the time direction on the twisted $d+1$ dimensional torus, is periodic only up a third root of unity $z \in \text Z_3 = \{1,\, \e^{2 \pi \i/3},\, \e^{4 \pi \i/3} \}$,
\be
L(\vec x + \vec e L ) = \e^{-2 \pi \i \,\vec e\cdot\vec k/3} \, L(\vec x) \, , \label{eq::PolyBCs}
\ee
corresponding to the non-trivial $\mathrm{SU}(3)$ center element that it picks up along the way from $\vec x$ inside the volume $V=L^d$ to $\vec x + \vec e L $ (i.e.~around the twisted torus along spatial directions). This leads to the  mirror-charge interpretation of the electric-flux ensembles \cite{ForcrandSmekal2002,Smekal2012}.

Here, we extend the notion of center electric fluxes to full Lattice QCD with dynamical Wilson fermions and explicitly construct ensembles with non-vanishing quark number $N_V\neq 0 \!\mod 3$ in a spatial subvolume $V$ of the lattice. Intuitively, the flux through its surface $S = \partial V$ from the charges inside $V$ must be compensated at all times by corresponding anti-charges in the spatial complement $\overline{V}$ of $V$ such that Gauss's law is satisfied and gauge-invariance therefore maintained. For example with $N_V = 1 \!\mod 3$ the one leftover quark in $V$ creates a flux line through $S$ which has to terminate in $\overline{V}$  on the other side. The only gauge-invariant way to achieve this is to introduce the corresponding anti-charge, e.g.~by an anti-quark  or a diquark in $\overline V$. We use this idea to construct our ensembles starting from a local $\mathbb{Z}_3$-Gauss law to fix the center charge inside $V$ by the center flux through $S$. Because the center charge is determined by the net quark number modulo three, fixing the center flux therefore fixes this net quark number in $V$. The basic ideas go back to Mack \cite{Mack1978}. The transfer-matrix formulation for the $\mathbb{Z}_3$-Gauss law in the pure gauge theory and its relation to 't Hooft's electric fluxes was elaborated by Borgs and Seiler in \cite{BorgsSeiler1983}. 
For a mathematical formulation of a hamiltonian approach based on operator algebras with fermions and a classification of the state space in terms of sectors of fixed center charges, see  \cite{KijowskiRudolph2002, KijowskiRudolph2002-2, KijowskiRudolph2005}.

The main problem addressed here, is how to combine these concepts with the dynamics of the system. This is subtle because, a priori, the presence of dynamical fermions does not leave the charged sectors invariant: There are processes which change the quark numbers in both volumes, a quark can hop from $V$ into $\overline{V}$ or a quark-anti-quark pair can be created between both volumes. It is this problem which we solve here. We present two main results. First, as we are going to see, we have to modify the dynamics at the surface $S$ in such a way that the charged sectors remain invariant, i.e.~such that thermal fluctuations cannot cause the system to leave a charged sector. Secondly, the corresponding path integral can be regarded as a generalization of 't~Hooft's electric flux ensembles. We only have to take discrete Fourier transforms over ensembles where timelike plaquettes bordering the volume $V$ are twisted. Contrary to 't Hooft's electric flux ensembles, we have to introduce twists between all Euclidean time slices separately. The fermion action remains unchanged, however.

The outline of our paper is as follows: In Sections \ref{sec::eft-ft} and \ref{sec::electrix_flux_ensembles}
we discuss the problem in the heavy-dense limit of QCD \cite{LangelageLottini2011, LangelageNeuman2014}. 
The effective Polyakov-loop theory of this limit can be approximated by a $\mathbb{Z}_3$-Potts model. The latter is equivalent to a flux-tube model \cite{Patel1984, DegrandDetar1983, BernardDegrand1994, CondellaDetar2000}. These effective theories are described in Section \ref{sec::eft-ft}. The flux-tube model exhibits a local $\mathbb{Z}_3$-law and therefore serves as a simple system for which we can construct the ensembles with fixed net quark numbers in a straightforward and illustrative manner. In Section \ref{sec::electrix_flux_ensembles}, we show that these ensembles are simply given by a discrete Fourier transform of  $\mathbb{Z}_3$-Potts models with twisted couplings at the surface of $V$. We also present numerical results for a one-dimensional chain and a three-dimensional lattice. In particular, the free energy differences between sectors of non-zero flux and zero flux through $S$ are computed. The last two sections, Section \ref{sec::towards_lattice_qcd} and \ref{sec::trans_const}, provide the extension to full Lattice QCD. We derive and justify our main results from two different approaches: the dualization of the fermion action (Section \ref{sec::towards_lattice_qcd}) as well as a more rigorous transfer-matrix formulation (Section \ref{sec::trans_const}). 
Our summary and conclusions are provided in Section \ref{sec::conclu}, and several appendices are added with further technical details.

\section{Effective Polyakov-loop theory and flux-tube model} 
\label{sec::eft-ft}

In this section we first consider the three-dimensional effective Polyakov-loop theory for heavy-dense QCD of Refs.~\cite{FrommLangelage2012,LangelageNeuman2014} for which, at leading order in the hopping expansion, the grand-canonical partition function reads (for one flavor of Wilson quarks),
\be
Z_\mathrm{eff} = \int \Big( \prod_i \d L_i \, J(L_i) \, Q(L_i)\Big) \,  \prod_{\langle i,j\rangle} \big(1+2\lambda \, \Re L_i L_j^* \big) \, , \label{eq::effact}
\ee
where $L_i = L(\vec x_i) $ is the Polyakov loop at spatial site $i$, $J(L)$ is the reduced Haar measure for $\mathrm{SU}(3)$, and $\langle i,j\rangle $ denotes the link between nearest-neighbor sites $i$ and $j = i+\hat{\mu}$, $\mu\in\{1,2,3\}$. The static fermion determinant \cite{DePietriFeo2007} factorizes and is included via one factor per site (and quark flavor) of
\be
 Q(L) = \big(1 + h L + h^2 L^* + h^3\big)^2 \big(1+\bar h L^* +\bar h^2 L + \bar h^3\big)^2, \label{eq::statDet}
\ee
where $h\equiv h(\mu) = \e^{(\mu-m)/T}$ and $\bar h \equiv \bar h (\mu) = h(-\mu)$.

The Polyakov-loop interaction is obtained from a character expansion of the original Wilson plaquette action, and subsequent integration of the spatial link variables, where $\lambda = \big(u \exp\{ 4 u^4 + 12 u^5 \pm \cdots\}\big)^{N_t}  $ can be computed to high orders from the strong-coupling expansion in terms of the fundamental character coefficient $u = \beta/18+\cdots $ and hence the gauge coupling $\beta $ \cite{MontvayMuenster1994}. In Eq.~(\ref{eq::effact}) we have used the form with partial resummation of nearest-neighbor fundamental Polyakov-loop interaction terms given in Ref.~\cite{LangelageLottini2011}.

\begin{figure}[t]
\includegraphics[width=0.8\linewidth]{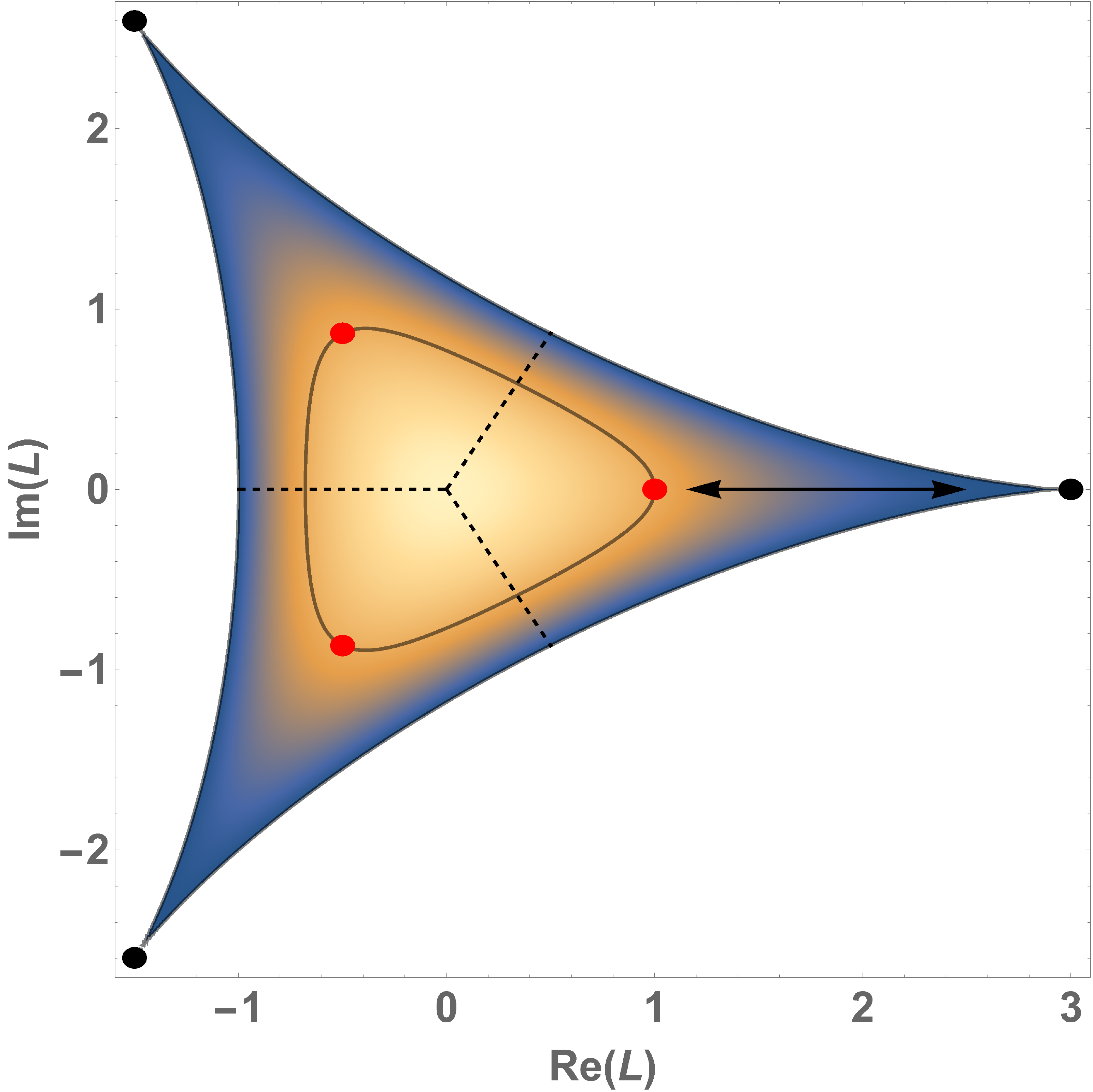}
\caption{Density plot of the reduced Haar measure with center elements (black dots) and third roots of unity (red dots) as set of midpoints for each of the three center sectors separated by dashed lines, and a double-headed arrow indicating the transition between weak and strong coupling, see text.}
\label{fig::redHaar}
\end{figure}

To emphasize the role of the electric $\mathbb Z_3$-center fluxes, we further reduce  the partition function in (\ref{eq::effact}) to a suitable three-states Potts model.
In order to reproduce the correct strong-coupling limit, we thereby implement the rules of group integration, as far as possible, in the $\text Z_3$-Potts model as well. To achieve this, we consider the moments $T_{m,n}$ with  $m, n = 0,1,\dots $ of the reduced Haar measure $J(L)$, as tabulated in Appendix B of Ref.~\cite{UhlmannMeinel2007},
\begin{equation*}
T_{m,n} = \int \d L\, J(L) \, L^m {L^*}^n\, .
\end{equation*}
For $m=n=0$ we have $T_{0,0}=1$ from the normalization of the Haar measure. Moreover, orthogonality of the group characters and center symmetry imply that $T_{m,n} =0 $ for all $m\not= n\!\mod 3$. The normalization of $L(W) = \tr W$, the fundamental character of $W \in \mathrm{SU}(3)$, is obtained with $m=n=1$,
\begin{equation*}   
(L,L)  = \int \d L\, J(L) \, |L|^2 = 1 \, .
\end{equation*}
In order to obtain a center-symmetry preserving approximation we first split
the integral of $|L|^2$ into a sum over the three center sectors and then use a midpoint 
definition for each of the three remaining integrals which amounts to replacing the Polyakov loops at $i$ and their integrations over the normalized Haar measure by
\begin{align}
L_i = \tr W_i \to  z_i \in \text Z_3 &= \left\{1,\, \e^{2 \pi \i/3},\, \e^{4 \pi \i/3} \right\} 
\, , \nonumber \\
\int \d L_i \, J(L_i) \, f(L_i) &\to \frac{1}{3} \sum_{z_i \in \text Z_3} f(z_i)\, . \label{eq::midpdef}
\end{align}
Each of the three roots of unity here thus represent a midpoint for the integration over the corresponding center sector of the reduced Haar measure, see Figure \ref{fig::redHaar}. Note that this is different from a naive center projection of the untraced straight Polyakov line $W_i \in \mathrm{SU}(3)$ at the spatial site $i$ which would lead to $ L_i = \tr W_i  \to 3 z_i $ and to hence fermionic site factors $Q(3 z_i)$ which do not reproduce 
the strong-coupling limit of heavy-dense QCD, see App.~\ref{sec::site-factors}, where the $\mathrm{SU}(3)$ group integrations of the Polyakov loops at the $N_s$ spatial sites in (\ref{eq::effact}) for $\lambda \to 0$, and $\bar h \to 0$ for $m \gg T$ (with $\mu>0$) yield,\footnote{Note that the local color-singlet three-quark states in the one-flavor theory are spin-$3/2$ baryons with a multiplicity of $4$, as counted by the degeneracy factor of the $h^3$ term.}
\be
Z_\mathrm{eff} \to (1+4 h^3 + h^6)^{N_s} %(1+4 \bar h^3 + \bar h^6)^{N_s}
\,. \label{eq::str-couplim}
\ee
In order to integrate an arbitrary class function $f(L)$ we would need to reproduce all moments $T_{m,n}$, however. This is obviously not possible with a simple midpoint definition. At least, the definition in (\ref{eq::midpdef}) correctly reproduces the orthogonality relations $T_{m,n}=0$ for $m\not= n\!\mod 3$, together with all non-vanishing moments up to $ m+n <4$, for which \cite{UhlmannMeinel2007}
\begin{equation*}
T_{m,n} %= \int \d L\, J(L) \, L^m {L^*}^n
= \left\{\begin{array}{ll}
0\, ,  &  m-n \not= 0 \!\mod 3 , \\
1\, , &  m-n = 0 \!\mod 3\, .
% & \; \& \;\; m+n < 4 \,. 
\end{array} \right.
\end{equation*}
The non-vanishing moments $T_{m,n} $ for $m+n \ge 4$ on the other hand grow factorially and this is of course not captured by the integration rule in (\ref{eq::midpdef}) which yields 1 for all $m=n\!\mod 3$.
The discrepancy first occurs at $m=n=2$ where $T_{2,2} =2$ for $\mathrm{SU}(3)$ \cite{UhlmannMeinel2007}. 
This moment counts the number of color singlets in the $ 3 \otimes 3 \otimes \bar 3 \otimes \bar 3$ product representation, of which we will keep only one instead of two. Therefore, in the strong-coupling limit $\lambda\to 0$, for example, we will miss some mesonic states such as tetra-quarks corresponding to $h^2\bar h^2$ in the site factors for the partition function of the effective theory in Eq.~(\ref{eq::effact}) with (\ref{eq::statDet}). These additional contributions are independent of chemical potential and moreover irrelevant when either $\bar h \to 0$ or $h\to 0$ for $m\gg T$ in the heavy-dense limit at either $\mu > 0 $ or $\mu<0$. They do therefore not affect the strong-coupling limit of heavy-dense QCD. For comparison we list the different site factors of the partition functions in the strong-coupling limit with group integration and $\text Z_3$ sums in Appendix \ref{sec::site-factors} explicitly.

Because Polyakov-loop distributions in the pure gauge gauge theory at finite gauge-coupling furthermore resemble those of the Haar measure with very high accuracy for all temperatures up to the deconfinement transition at $T_c$ \cite{SmithDumitru2013,EndrodiGattringer2014}, the midpoint definition (\ref{eq::midpdef}) remains equally well justified,  in fact, for the corresponding low-order moments $\langle L^m {L^*}^n\rangle $ with  $m+n <4 $ of the Polyakov-loop in the pure gauge theory throughout the confined phase.

With the $\text Z_3$-symmetric midpoint definition (\ref{eq::midpdef}), the grand canonical partition function in Equation (\ref{eq::effact}) thus becomes
\begin{equation}
    \label{eq::effPotts}
    Z_\mathrm{eff} = \frac{1}{3^{N_s}} \sum_{\{z_i \in \text Z_3\}} \prod_{\langle i,j\rangle} \left(1+2\lambda \, \Re z_i z_j^* \right)\,\prod_i Q(z_i)\, ,
\end{equation}
and one easily verifies explicitly that the strong-coupling limit (\ref{eq::str-couplim}) is reproduced correctly. 

A straightforward calculation along the lines of Ref.~\cite{CondellaDetar2000} shows that the effective Potts model for heavy-dense QCD in Equation (\ref{eq::effPotts}) is equivalent to a variant of the flux-tube model presented therein. Our variant here is defined by independent site-occupation numbers 
\begin{equation*}
n_{i,s} \in \{0,\dots,3\} \; \mbox{and} \;\; \bar n_{i,s} \in \{0,\dots,3\}
\end{equation*}
for quarks and anti-quarks with spin $s \in \{ \uparrow, \downarrow \}$, connected with fluxes represented by link variables
\begin{equation*}
l_{\langle i,j\rangle} \in \{-1,0,1\} \, . 
\end{equation*}
The energy of combined (anti-)quark and flux-line configurations $\{n,l\}$, with mass $m$, string tension $\sigma$ and lattice spacing $a$, is 
\be
H(n,l) = \sum_{\langle i,j\rangle} \sigma a |l_{\langle i,j\rangle} | + \sum_{i,s} m ( n_{i,s} + \bar n_{i,s}) \, , \label{eq::ftHdef}
 \ee
 and the allowed {\em physical} configurations are restricted by a $\mathbb Z_3$-Gauss law on the lattice, relating the divergence $\phi_i$ of the fluxes to the net quark number $q_i$ at site $i$, 
 \be
  \sum_s ( n_{i,s} - \bar n_{i,s} ) - \sum_{j\sim i}  l_{\langle i,j\rangle }\equiv q_i-\phi_i  = 0 \!\mod 3\, . \label{eq::lGauss}
 \ee
 Here, the sum over $j\sim i$ sums the flux lines from all links that connect site $i$ with one of its nearest neighbours at $j\sim i$ (forward and backward), and we generally have $l_{\langle j,i\rangle} = - l_{\langle i,j\rangle}$, so that the second sum on the left in (\ref{eq::lGauss}) is the lattice divergence of the link variables and yields the total flux emanating from site $i$. The factor $\sigma a$ is the energy of the smallest possible flux tube, i.e. a flux tube of length $a$. We refer to $\sigma a$ as the string energy unit. In addition, we define the total net quark number of a configuration as $q = \sum_{i}q_i$.

 The grand-canonical partition function then agrees %, up to a normalization factor $\mathcal N$,
 with that of our effective Potts model (\ref{eq::effPotts}) for heavy-dense QCD, i.e.,
\be
Z_\mathrm{eff}(T, \mu) \propto %\mathcal{N} \!\!
\!\sum_{\{n,l\}_\mathrm{phys.}} \!\!\exp\Big\{\! - \beta\Big( H(n,l) - \mu q  \Big) \Big\} \, ,  \label{eq::effFTM}
 \ee
 with $\beta = 1/T$ and $\lambda = e^{-\beta \sigma a}$.
 
 The essential difference between our flux-tube model for heavy-dense QCD in Equation (\ref{eq::effFTM}) and the original one in Ref.~\cite{CondellaDetar2000} is the treatment of quark and anti-quark occupation numbers. Whereas the static quark determinant for heavy-dense QCD in (\ref{eq::effPotts}) requires independent occupation numbers $n_{i,s}$ and $\bar n_{i,s} \in \{ 0,\dots 3\}$
 for quarks and anti-quarks (and spin), the model of Ref.~\cite{CondellaDetar2000} assumes common occupation numbers $n_{i,s} \in \{ -3,\dots 3\}$ for quarks and anti-quarks, which would lead to common site factors for quarks and anti-quarks  in the fermion determinant, see Appendix~\ref{sec::site-factors}.  While the resulting multiplicities of three and six quark and anti-quark states agree with those of the strong-coupling limit of the effective theory in (\ref{eq::effact}), this would lead to a multiplicity of two for the quark-anti-quark states corresponding to $ h \bar h $, in the strong-coupling limit, $\lambda \to 0 $ resp.~$\sigma a\to \infty$, see Appendix \ref{sec::site-factors}.
 With the spin-statistics theorem intact, this would imply two scalar mesons in the one-flavor theory where there should be one spin-0 and one spin-1 meson with the site factors (\ref{eq::statDet}) of the static fermion determinant.
 
Beyond the strong-coupling limit, when the string energy unit is finite, quarks and anti-quarks or diquarks can be separated across different lattice sites as long as they remain connected by corresponding flux lines, consistent with the $\mathbb Z_3$-Gauss's law.

\begin{figure}[t]
  \vspace*{.4cm}
\includegraphics[width=0.8\linewidth]{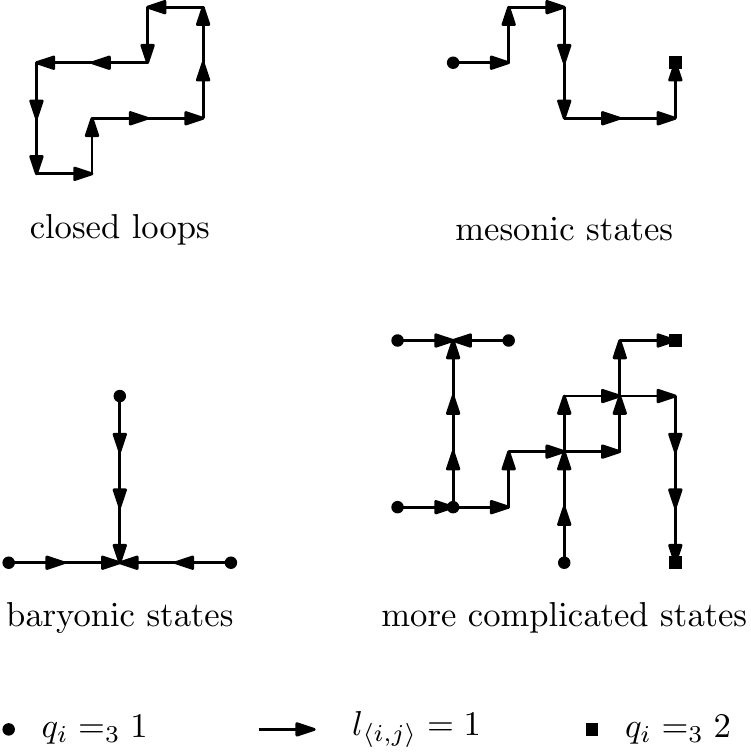}
\caption{Examples of {\em physical} (anti-)quark and flux-line configurations in $d=2$ dimensions that are allowed by the $\mathbb Z_3$-Gauss's law in Equation (\ref{eq::lGauss}).}
\label{fig::example-states}
\end{figure}

Before we continue to introduce interfaces corresponding to 't~Hooft's twisted boundary conditions (\ref{eq::PolyBCs}) and center fluxes according to the electric fluxes in (\ref{eq::elf}), in the next section, we note that the effective $\text Z_3$-Potts model of heavy-dense QCD in (\ref{eq::effPotts}) can be rewritten exactly, in a perhaps more conventional form,
\begin{equation}
    \label{eq::effPottsexp}
     Z_\mathrm{eff} =  \mathcal N \sum_{\{z_i \in \text Z_3\}} \exp\bigg\{\sum_{\langle i,j\rangle}  2\gamma \, \Re z_i z_j^*  \bigg\}\,\prod_i Q(z_i)
\end{equation}
with the relations
\begin{equation}
\gamma = \frac{1}{3} \ln\left(\frac{1+2\lambda}{1-\lambda} \right) = \frac{1}{3} \ln\left( 1+ \frac{3}{e^{\beta \sigma a} - 1} \right) \, ,   \label{eq::Potts_gamma}
\end{equation}
and normalization $\mathcal N =  \big[(1+2\lambda)(1-\lambda)^{2}\big]^{d N_s/3}/3^{N_s}$.

The effective Potts model has inherited the Roberge-Weiss symmetry of QCD in form of its global $\text Z_3$ symmetry, and so does its equivalent flux-tube model; i.e.~at imaginary chemical potential $\mu = \i \theta T$,
\begin{equation*}
Z_\mathrm{eff}(T,\i \theta T) \equiv Z^I_\mathrm{eff}(\theta) =Z^I_\mathrm{eff}(\theta + 2\pi/3) \,.
\end{equation*}
Without interfaces or any non-periodic boundary conditions, the corresponding canonical partition functions $Z_\mathrm{eff}^c(T,N_q)$ therefore still vanish whenever the total quark number $N_q = \sum_i q_i \not= 0 \!\mod 3$. Using total baryon number $N_B $ instead,  with $N_q = 3 N_B$, we can therefore write
\begin{equation*}
 Z^I_\mathrm{eff}(\theta) = \sum_{N_B =- N_B^\mathrm{sat}}^{N_B^\mathrm{sat}}  \e^{\i 3N_B \theta}\, Z_\mathrm{eff}^c(T,3N_B) \,,
\end{equation*}
where saturation occurs at $N_B^\mathrm{sat}  = N_q^\mathrm{sat}/3 = 2 N_f N_s$ on a finite lattice with $N_f$ flavors and $N_s$ sites. 

This is therefore a good starting point to introduce interfaces and electric center fluxes in our flux-tube model representation of the effective theory for heavy-dense QCD in the next section.

\section{Electric flux ensembles}
\label{sec::electrix_flux_ensembles}

As we have seen in the previous subsection, the flux-tube model (\ref{eq::effFTM}) provides an illustrative realization of the center-electric fluxes between neighboring sites. This can directly be used to define the total flux $\phi_S$ of a particular configuration through an arbitrary surface $S$ by
\begin{equation*}
\phi_{S} = \sum_{\langle i,j\rangle \in S^\ast} l_{\langle i,j\rangle}\,,
\end{equation*}
where the sum runs over the (in three dimensions dual) stack of links that intersect $S$ in the direction of its orientation. 
The corresponding interface $S^*$ inherits the orientation of $S$, i.e., we have $\phi_{-S} = -\phi_{S}$, and $-S^\ast$ is the interface that contains the same links as $S^\ast$ but with reversed directions. In order to convert the sum into one over all forward-directed links as before, we define the local orientation of $S$,
\begin{equation*}
    s_{\langle i,j\rangle} = 
    \begin{cases} +1, & \langle i,j\rangle \in S^*\\
     -1, & \langle j,i\rangle \in S^* \\
     \phantom{+}0, & \mbox{otherwise}
     \end{cases}
\end{equation*}
which allows us to write 
\begin{equation}
\phi_{S} = \sum_{\langle i,j\rangle}  s_{\langle i,j\rangle} l_{\langle i,j\rangle}\,, \label{eq::fuxdef}
\end{equation}
where the sum over forward links with $j=i+\hat\mu$ is implied.

As a warm-up exercise, we first consider the pure $\mathrm{SU}(3)$-gauge theory in which the fermion determinant is unity, and we only need to consider the nearest-neighbor interaction of Potts spins. Equivalently, with $ m\to \infty $ in the flux-tube model, we are left with pure flux-tube configurations $\{n=0,l\}$ obeying the local $\mathbb{Z}_3$-Gauss law. 
In order to introduce a $\mathbb Z_3$-interface $S^\ast$, we first define the Kronecker delta modulo three,
\begin{equation}
    \delta_3(\phi)  = \frac{1}{3}\sum_{z\in\mathrm{Z}_3} z^{\phi}  \, . \label{eq::Kdm3}
\end{equation}
With this we can fix the electric flux through $S$ to a certain value $e \in\mathbb{Z}_3$ by restricting the partition function of the flux-tube model to configurations with $\phi_{S} = e\!\mod 3$, for which we introduce the shorthand notation $\phi_S =_3 \, e$ (where the symbol `$=_3$' denotes the modulo three part of any integer in the following),
\begin{equation}
    \label{eq::efe_pure_ftm}
    \begin{split}
        Z(\phi_{S} &=_3 e) =\\
        &\sum_{\{l\}_{\mathrm{phys.}}}\delta_3(\phi_{S} - e) \, \exp\big\{-\beta\sum_{\langle i, j\rangle}\sigma a\,|l_{\langle i, j\rangle}|\big\}\, .
    \end{split}
\end{equation}
The derivation of the equivalent Potts-model representation of this restricted partition function proceeds analogous to that without such restriction,  and yields
\begin{equation}
    \label{eq::efe_pure_fourier}
Z(\phi_S =_3 e) \propto \frac{1}{3}\sum_{k\in\mathbb{Z}_3}\e^{-\frac{2\pi\i}{3}ke} \, Z_{S}(k)
\end{equation}
with suitably \emph{twisted} ensembles 
\begin{align*}
\begin{split}
Z_{S}(k) = \sum_{\{z_i\in \text Z_3\}}& \exp\bigg\{\sum_{\langle i,j\rangle}2\gamma\, \Re \left(z^{-s_{\langle i,j\rangle}k}\,
z_i z_{j}^*\right)  \bigg\}\,,
\end{split}
\end{align*}
and $z = \e^{2\pi\i/3}$. These twisted ensembles differ from the standard Potts-model partition function in  (\ref{eq::effPottsexp}) by factors of $z^{\mp k} $ in all nearest-neighbor couplings between the spins with $\langle i,j\rangle \in \pm S^*$ which is precisely how one implements the (cyclic) interface $S^*$ in the Potts model \cite{Smekal2012}. 

The discrete Fourier transform (\ref{eq::efe_pure_fourier}) over these twisted Potts-model ensembles $Z_{S}(k)$ is therefore equivalent to a flux-tube model in which the flux through a surface $S$ is fixed to $\phi_S=e\!\!\mod 3$.  When the surface is closed,  $S=\partial V $, the interface can be removed from  $Z_{S}(k)$ by the substitution $z_i \to z^k z_i $ for all spins inside $V$, just as closed center-vortex sheets can be removed without free-energy cost in the pure gauge theory. When $S$ extends across the torus, it introduces a seam with a cyclic shift which can be straightened (but not removed) and hence corresponds to introducing 't Hooft's twisted boundary conditions in the pure gauge theory. The relation (\ref{eq::efe_pure_fourier}) between the twisted $Z_{S}(k)$ and the electric-flux ensemble $Z(e)$ is then analogous to that between 't Hooft's twists and the electric fluxes in Eq.~(\ref{eq::elf}). The only difference is that the one-dimensional $\mathbb Z_3$ Fourier transform in Eq.~(\ref{eq::efe_pure_fourier}) only fixes the center flux perpendicular to the plane of the straightened $S$,  but not the electric fluxes in the in-plane directions where the boundary conditions remain periodic, whereas in Eq.~(\ref{eq::elf}) the electric fluxes are fixed in all $d$ spatial directions.

\begin{figure}
\centering
\includegraphics[scale=1.0]{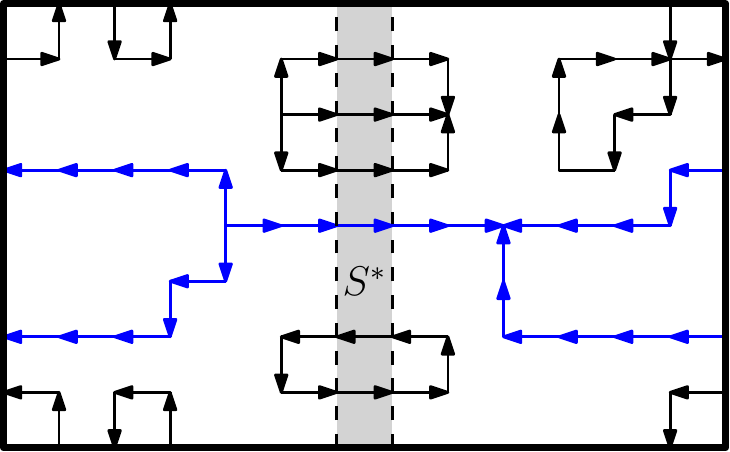}
\caption{Example of a \emph{physical} flux-line configurations with $\phi_{S} = 1\mod 3$ in the absence of quarks and anti-quarks.}
\label{fig::example-states-interface}
\end{figure}

An example of a flux-line configuration with $\phi_S = 1 \!\mod 3$ in $d=2$ dimensions is shown in Figure~\ref{fig::example-states-interface}.
A non-zero flux  (with $l_{\langle i, i+\hat\mu \rangle} = \pm 1$) is indicated by a forward arrow (in the direction from $i$ to $j=i+\hat\mu$) for  $l_{\langle i, j\rangle} = + 1$, and by a backward arrow (from $j$ to $i$) for  $l_{\langle i, j\rangle} = - 1$. 
No arrow means no flux, $l_{\langle i, j\rangle} = 0$, in this picture. A physical configuration $\{n=0,l\}$ is made of closed flux-tube loops due to the $\mathbb{Z}_3$-Gauss law. These loops consist of lines of arrows along the links of the lattice which can be connected by vertices at the sites of the lattice where the difference of incoming and outgoing arrows must always be a multiple of three, due to the $\mathbb{Z}_3$-Gauss law. Loops constructed in this way can not create a non-zero flux $\phi_S \neq 0 \! \mod 3$ without winding around the lattice. Hence, physical configurations with $\phi_S \neq 0\! \mod 3$ are necessarily made of loops winding around the lattice in the direction perpendicular to $S$ (such as the blue \emph{torus cluster} in Fig.~\ref{fig::example-states-interface}).  

After having established the analogy between the pure electric-flux ensembles in the flux-tube model and 't~Hooft's electric fluxes in the pure gauge theory, as the next step we now reintroduce the static quarks and anti-quarks of heavy-dense QCD. The flux tubes are then no-longer forced to form closed loops but can start and end at the sources and sinks that arise from the static quarks and anti-quarks or diquarks, as long as this is compatible with the $\mathbb{Z}_3$-Gauss law in (\ref{eq::lGauss}). At the same time, the total electric flux (mod $3$) through a closed surface $S=\partial V$ need not be zero anymore, but determines the net number of center charges in $V$, and the corresponding interface can in general not be removed by a transformation of variables.

We can define ensembles with $\phi_S \neq 0 \!\mod 3$ but this does not require the existence of a winding loop. For example, we can have a quark on one side of $S$ connected to an anti-quark on the other side by a flux line through the interface $S^\ast$ which generates a flux $\phi_S \neq 0\!\mod 3$. In order to see explicitly how the $\mathbb{Z}_3$-Gauss law can be used to restrict the net quark number in a subvolume $V$ to a certain value modulo three, consider some volume $V$ comprising a certain subset of all lattice sites, with surface $S=\partial V$ separating $V$ from its complement $\overline V$. Then, $ S^\ast $ consists of all links connecting $V$ with its complement $\overline{V}$. The links of $S^\ast$ always start at a site in $V$ and end at the adjacent site in $\overline{V}$. The local $\mathbb{Z}_3$-Gauss law in (\ref{eq::lGauss}) then establishes the equivalence, up to a multiple of three, of the flux $\phi_S$  through $S^\ast$, defined in Eq.~(\ref{eq::fuxdef}), for the closed surface $S=\partial V$ and the total charge $q_V$ in $V$, inside $S$,
\begin{equation}
    \label{eq::global_gauss_ftm}
    \phi_S = q_V\!\!\mod 3\,,\quad\mbox{where} \;\; q_V = \sum_{i\in V}q_i\,.
\end{equation}
Obviously, if $S^\ast = \varnothing $, then all sites of the lattice are contained in either $V$ or $\overline V$ and $\phi_S = 0$.
In particular, this confirms that physical configurations can only have multiples of three as their total net quark number $q$.

With Equation (\ref{eq::global_gauss_ftm}), an ensemble with fixed net quark number modulo three in $V$ is thus the same as an ensemble with fixed flux through $S=\partial V$,
\begin{equation*}
    Z(q_V =_3 e) = \sum_{\{n,l\}_{\mathrm{phys.}}}
    \delta_3(\phi_{S} - e) \; \e^{-\beta (H- \mu q)}\, .
\end{equation*}
This is analogous to Equation (\ref{eq::efe_pure_ftm}) but now with the full flux-tube model Hamiltonian (\ref{eq::ftHdef}) for finite quark mass $m < \infty $ and quark chemical potential $\mu$ included. The other difference here is that set of links in $S^*$ is now dual to the closed surface $S=\partial V$ which we could remove at no free-energy cost for $m\to\infty $ without the quarks before.
The same steps that we used to derive Eq.~(\ref{eq::efe_pure_fourier}) then again yield
\begin{equation}
    \label{eq::efe_quark_fourier}
    Z(q_V =_3 e) \propto \frac{1}{3}\sum_{k\in\mathbb{Z}_3}\e^{-\frac{2\pi\i}{3}ke}\, Z_S(k)
\end{equation}
but now with $S^\ast$ as the coboundary of $V$ and, importantly, with the same fermionic site factors  of heavy-dense QCD as in (\ref{eq::effPottsexp}), reintroduced unchanged in the definition of the ensembles  (with $z = \e^{2\pi\i/3}$ as before) 
\begin{align}
Z_{S}(k) &= \label{eq::efe_effPotts}\\ 
&\hskip -.6cm  \sum_{\{z_i\in \text Z_3\}} \exp\bigg\{\sum_{\langle i,j\rangle}2\gamma\, \Re \left(z^{-s_{\langle i,j\rangle}k}\,
z_i z_{j}^*\right)  \bigg\}\,\prod_i Q(z_i)\,, \nonumber
\end{align}
here. In this way, the basic structure in 't~Hooft's electric flux ensembles, as discrete Fourier transforms
of twisted ensembles, directly translates to the flux-tube model with quarks and anti-quarks where it allows to fix the net quark number in a subvolume $V$ (modulo three). 

To define the (Landau) free energy difference $\Delta F$ of an ensemble with such an electric flux $\phi_S=_3 e$ through $S$ relative to one with $\phi_S =_3 0$,  it is convenient to first divide the lattice in half along one direction, say we consider lattices of even extension $L_1$ and choose $V$ as all lattice sites $i$ with $i_1 \in  \{0,\ldots,L_1/2 -1\}$. 
We then define the electric-flux free energy as the large $L_1 $ limit of $\Delta F$ from
\begin{equation}
    \label{eq::def_free_energy}
    \Delta F_{\infty} = \lim_{L_1\rightarrow\infty} \left[-\frac{1}{\beta}\ln\left(\frac{Z(q_V =_3 1)}{Z(q_V =_3 0)}\right)\right]\,,
\end{equation}
where the same temperature, chemical potential and perpendicular extensions $L_2$, $L_3$ are assumed in both ensembles. This definition is analogous to that used for 't~Hooft's string tension \cite{tHooft1979,BorgsSeiler1983}, with  the obvious difference that the factor $L_1^{-1}$ is missing, reflecting the fact that it describes the total electric-flux free energy rather than that per length,  
as the string tension in the pure gauge theory does.
With the inclusion of (heavy) quarks and anti-quarks,
the flux string will break eventually, and localized quark-anti-quark structures, which do not scale with $L_1$ asymptotically, will create the required flux through the surface $S$ in the $L_1\to\infty$ limit. 

To make this argument more precise it is useful to study  Equation (\ref{eq::def_free_energy}) in the limits of very high and very low temperatures, where we can illustrate under reasonable assumptions how $\Delta F_{\infty}$ remains finite for $L_1\to\infty  $ in either case. We will explicitly verify this numerically for a one-dimensional chain and a three-dimensional lattice in the next subsections below. 

At high temperatures, for $\beta > 0$ sufficiently small, the free energy $\Delta F$ can be expanded in powers of $\beta$ such that
\begin{equation}
    \Delta F = -\frac{1}{\beta}\ln\left(\frac{M_1}{M_0}\right) + \frac{\alpha_1}{M_1}-\frac{\alpha_0}{M_0} +\mathcal{O}(\beta) \label{eq::highTDF}
\end{equation}
with $M_e = Z(q_V =_3 e, \beta = 0)$ and 
\begin{equation*}
    \alpha_e= \sum_{\{n,l\}_{\mathrm{phys.}}}\!\delta_3(\phi_S- e)\,\big(H-\mu q\big)\,.
\end{equation*}
The ratio $M_1/M_0$ can be computed exactly from the equivalent Potts-model formulation:
\begin{equation*}
    \frac{M_1}{M_0} = \frac{1+4^{-4N_s} - 2\cdot 4^{-2N_s}}{1+4\cdot 4^{-4N_s} + 4\cdot 4^{-2N_s} }\,.
\end{equation*}
On any finite lattice with $N_s$ sites,
this ratio is obviously different from unity, but it approaches unity in the infinite-volume limit. Therefore,
    $\ln({M_1}/{M_0}) \to  0 $
in this limit and the subsequent high temperature limit $\beta\to 0$ of $\Delta F$ thus exists. Moreover, we find
\begin{equation*}
    \begin{split}
        \frac{\alpha_e}{M_e} = 2\sigma a\, N_s  + \frac{6m \, N_s + \mathcal{O}(N_s\cdot 4^{-2N_s})}{1+\mathcal{O}(4^{-2N_s})}\, ,
    \end{split}
\end{equation*}
independent of the flux $e$,
such that the terms of order $\beta^0$ in (\ref{eq::highTDF}) in total vanish for $N_s\rightarrow \infty$ as well. 
In this order of limits,
$\Delta F_\infty$ thus indeed tends to zero, i.e.~in the thermodynamic limit ($N_s\to\infty$, taken first), the free-energy difference of both ensembles therefore  vanishes as $\beta\to 0$ at very high temperatures.

Before we discuss the low-temperature expansion next, we first note that it has been important above to define the electric-flux free energy relative to the ensemble where the flux through $S$ must vanish, which is not the same as assuming there is no interface in the first place. In fact, had we used the unrestricted partition function $Z_{\mathrm{eff}}$   from (\ref{eq::effPottsexp}), without interface, rather than $Z(q_V =_3 0)$ for the normalization of the electric-flux free energy in  (\ref{eq::def_free_energy}), we would have obtained a divergent infinite-temperature limit, instead. This is simply because non-trivial twists are generally suppressed at high temperatures which, in turn, implies that $Z(q_V =_3 0)$ tends to $1/3$ of the $k=0$ term in the $\mathrm{Z}_3$-Fourier transform (\ref{eq::efe_quark_fourier}). The latter corresponds to the full partition function $Z_{\mathrm{eff}}$, on the other hand. 
Hence, the corresponding coefficient of the leading term of order  $1/\beta$ in the high-temperature expansion (\ref{eq::highTDF}) would then have a finite limit, $\ln(M_1/M_{\mathrm{eff}}) \to  - \ln{3}$ for $N_s\to\infty$.

To analyse our electric-flux free energies in the low temperature limit, consider the energies
\begin{equation*}
\Delta_e(n,l) = \big(H(n,l) - \mu q(n)\big)\Big|_{\mathrm{phys.},\, q_V=_3 e} 
\end{equation*}
of physical flux-tube model states with $\phi_S=_3 q_V =_3 e$.
We label these in increasing order  starting from the lowest possible one  $\Delta_e^{(0)} < \Delta_e^{(1)} < \cdots$. Sorting the sums in both partition functions correspondingly, the electric-flux free energy can be expanded in $1/\beta$ as follows,
\begin{align}
    \label{eq::approx_free_energy}
    \Delta F &= (\Delta_1^{(0)}-\Delta^{(0)}_0)\\
    &\hskip .4cm-\frac{1}{\beta}\ln\left(\frac{N_1^{(0)} + N_1^{(1)}\e^{-\beta (\Delta_1^{(1)}-\Delta_1^{(0)})}+\ldots}{N^{(0)}_0 + N^{(1)}_0\e^{-\beta (\Delta^{(1)}_0-\Delta^{(0)}_0)}+\ldots}\right)\, , \nonumber
\end{align}
where the $N_e^{(i)}$ are the multiplicities of the corresponding energy levels. If the terms in this expansion converge in the $L_1\to\infty $ limit, as they did in the high temperature expansion, we can therefore conclude that  
\begin{equation}
    \label{eq::low_temp_free_energy}
 \Delta F_{\infty} =  \lim_{L_1\to\infty} \Big(
      \Delta_1^{(0)}-\Delta^{(0)}_0 \Big) 
      + \mathcal O(1/\beta)
%      \, ,\;\; \mbox{for}\;\; \beta\rightarrow\infty\, . 
 \end{equation}
in the zero-temperature limit. Unlike the high-temper\-ature limit above, where the relevant thermodynamic limit 
$N_s\to\infty$ is independent of the aspect ratio, the zero-temperature result will depend on the perpendicular sizes in more than one spatial dimensions, however.
Before we discuss this, we first study a one-dimensional chain where there are no perpendicular extensions of the interface and this additional subtlety does not arise. This therefore serves as the arguably most simple system to illustrate the novel features of our ensembles consisting of subvolumes with fixed quark numbers and non-trivial center fluxes across the boundaries between them.

\begin{figure}
    \centering
    \includegraphics[scale=0.70]{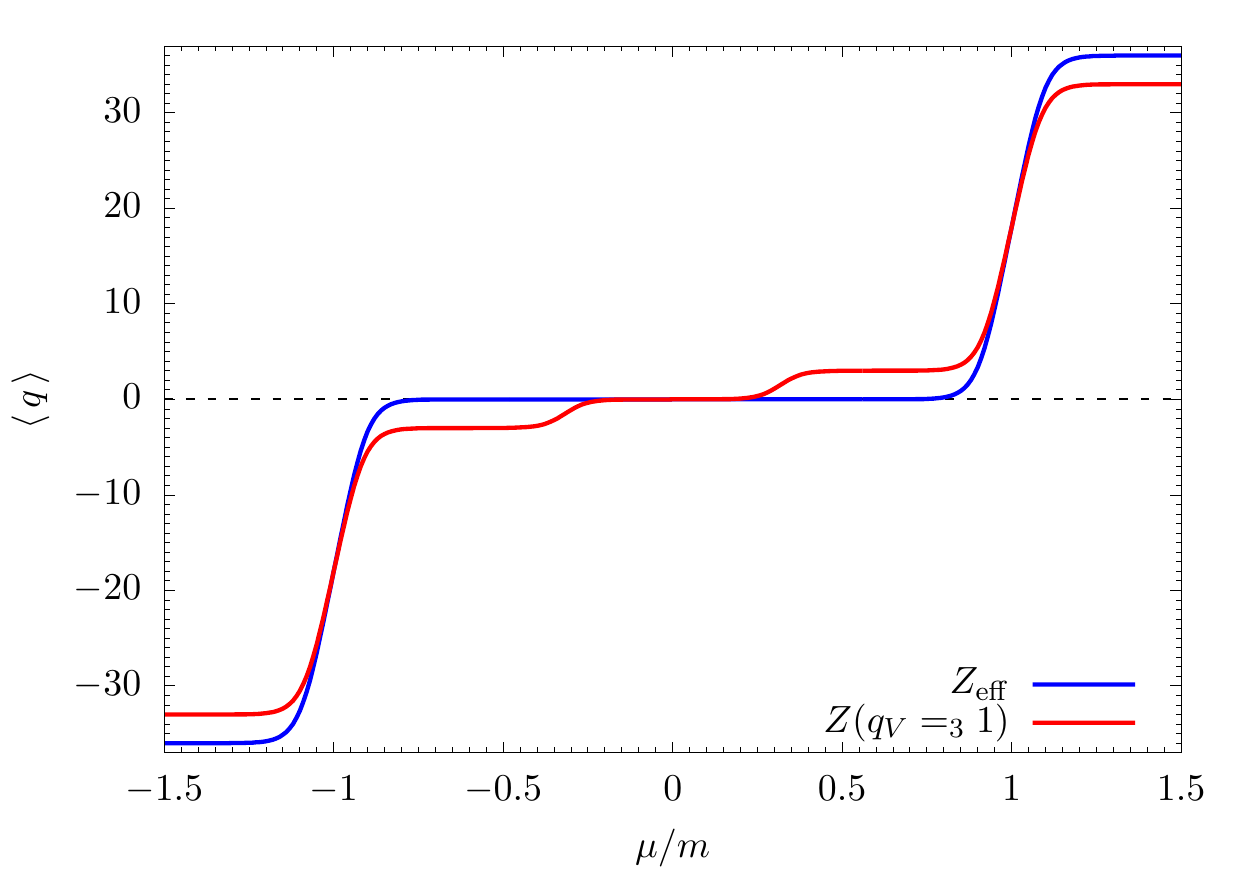}
    \caption{Total net quark number $\langle q \rangle$ of the two ensembles $Z_{\mathrm{eff}}$ and $Z(q_V =_3 1)$ for $L=6$, $\sigma a/m = 0.3$ and $T/m = 0.1$.}
    \label{fig::plot_fnf_comp}
\end{figure}

\begin{figure*}
%\begin{subfigure}{0.33\linewidth}
\centering
\includegraphics[scale=0.46]{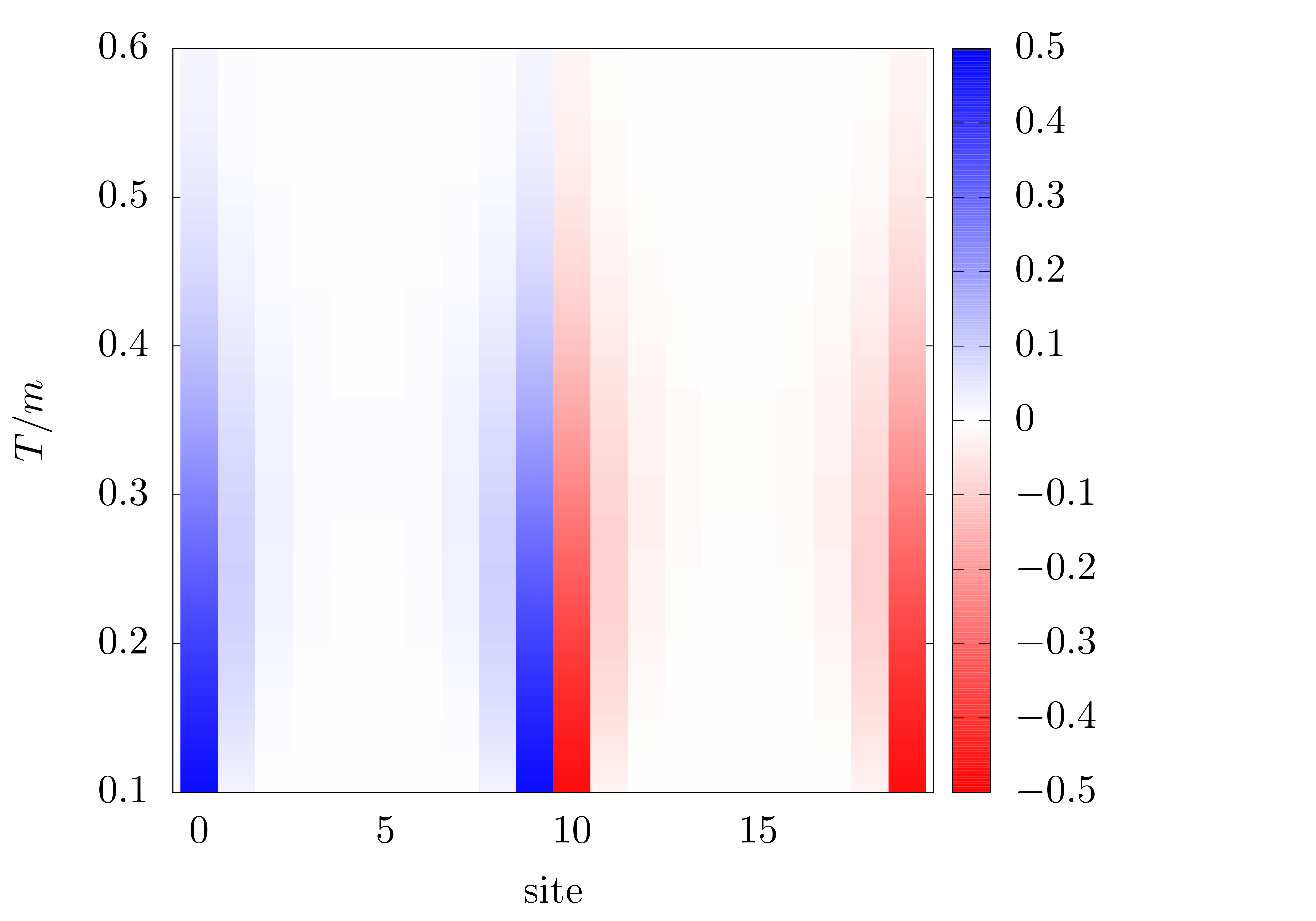}
\includegraphics[scale=0.46]{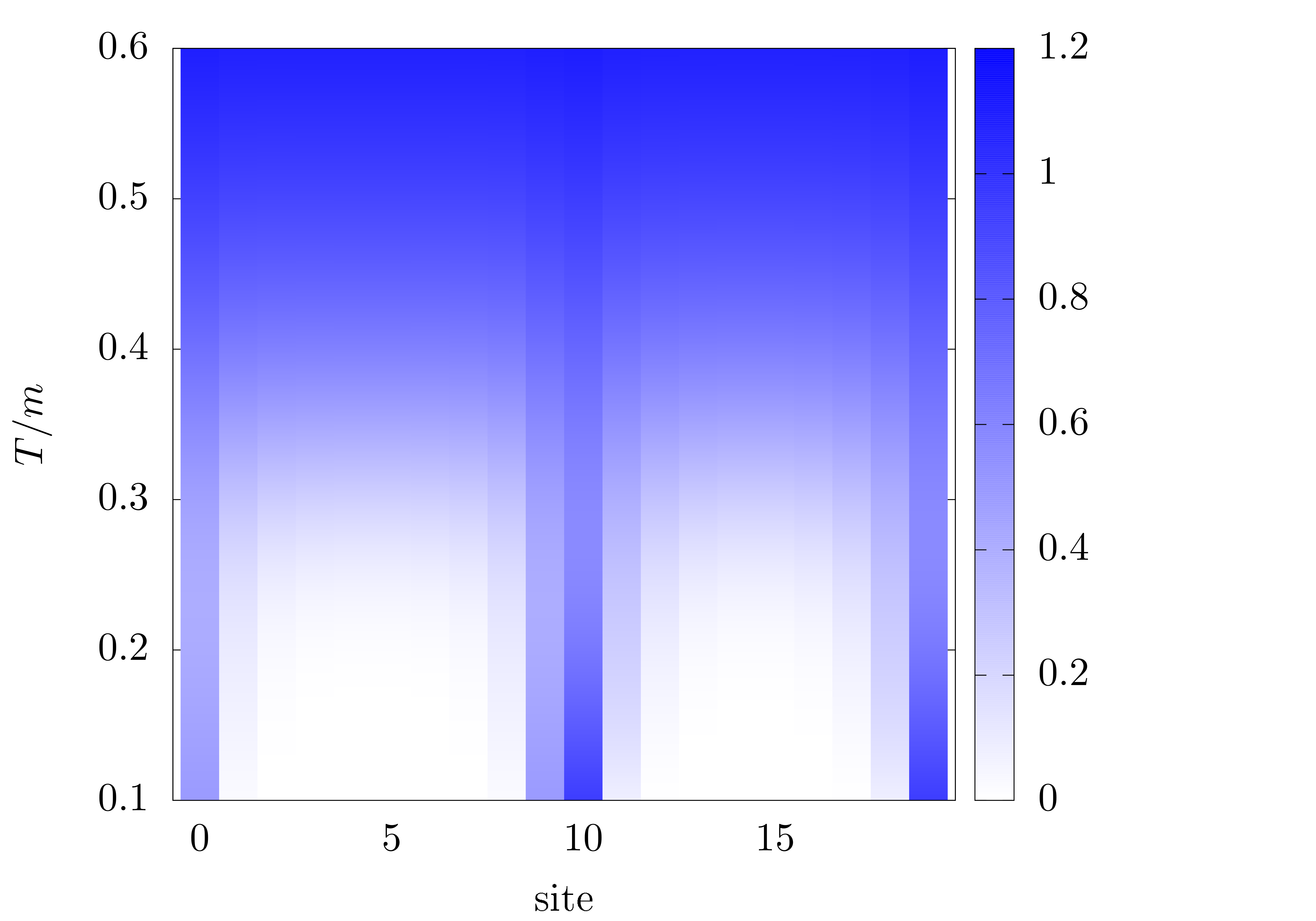}
\includegraphics[scale=0.46]{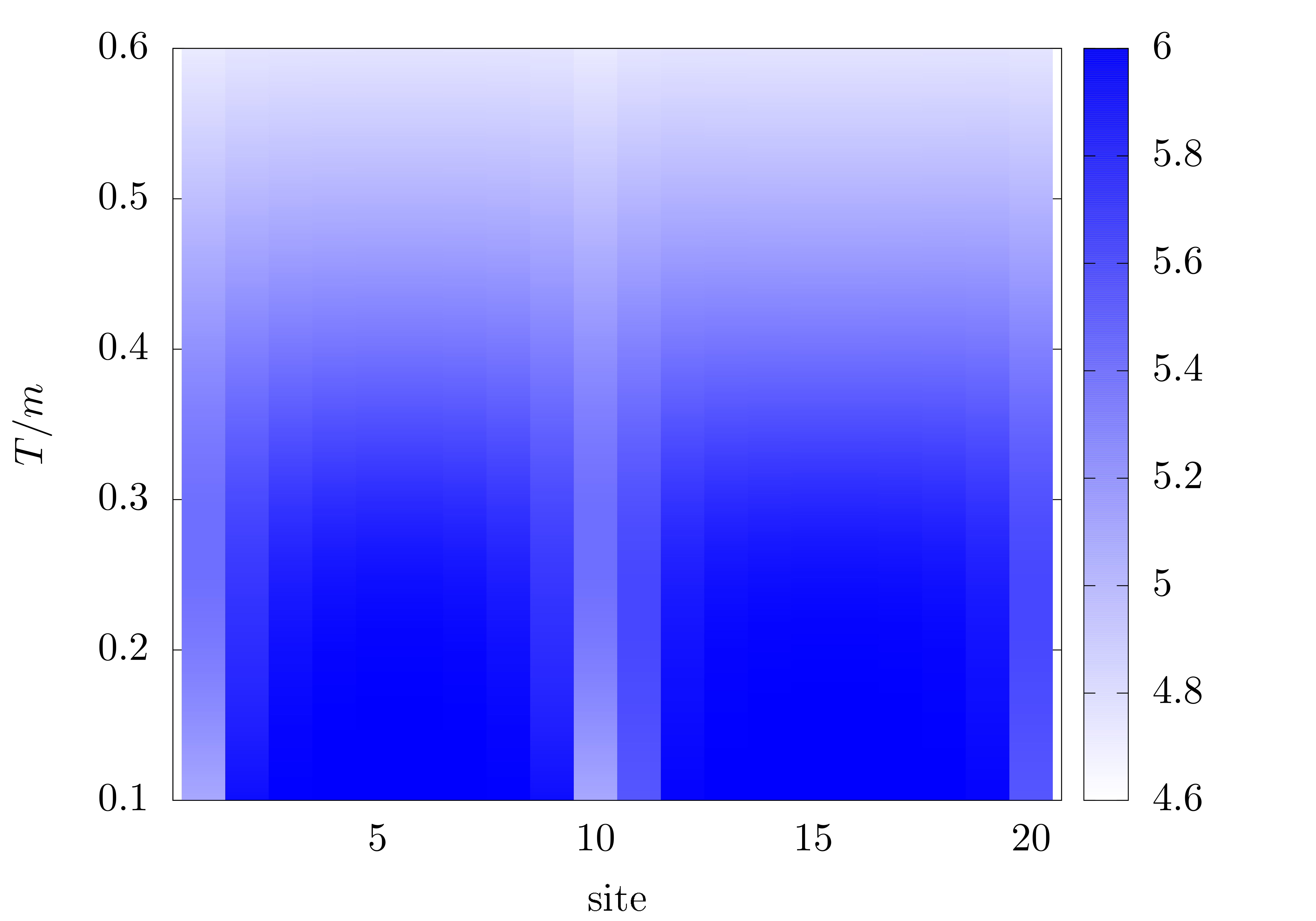}

\hspace*{-.4cm}
\parbox{0.25\linewidth}{(a)\hskip .2cm $\mu/m = 0$ (mesonic)} \hspace{1.2cm}
\parbox{0.25\linewidth}{(b)\hskip .2cm $\mu/m = 0.5$ (baryonic)} \hspace{1.4cm}
\parbox{0.25\linewidth}{(c)\hskip .2cm $\mu/m = 1.5$ (saturation)}

\caption{Net quark number $\langle q_i\rangle$ at a site $i$ of a chain with $L=20$ in the ensemble $Z(q_V =_3 1)$ with $\sigma a/m = 0.3$. The plots show the net quark number depending on the temperature $T/m$. The color coding represents the value of $\langle q_i \rangle$.}
\label{fig::plot_qnd_1d}
\end{figure*}

\subsection{One-Dimensional Chain}

As a first example we consider a one-dimensional chain of even length $L$ divided into two halfs, with sites $i\in  \{0,\ldots,L/2 -1\}$ in 
 $V$ and  sites $i \in  \{L/2,\ldots,L -1\}$ in $\overline{V}$, and $L\to \infty $ in the infinite volume limit. This implies that $S^*$ consists of only two links, the link from site $L/2-1$ to site $L/2$ as well as that from site $0$ backwards to site $L-1$, which intersect the surface $S=\partial V$.
 
 For the computations of the electric flux ensembles we use the equivalent Potts-model formulation from Equations (\ref{eq::efe_quark_fourier}) and (\ref{eq::efe_effPotts}). 
 For sufficiently small lengths $L$ we can compute the partition functions analytically. For larger $L$, where a direct evaluation of the sums 
 over spin configurations is no-longer feasible, we use the well-known transfer-matrix approach for one-dimensional spin systems as described in standard texts, e.g., see Refs.~\cite{Kadanoff2000} and \cite{Schwabl2006}. In the transfer-matrix approach, the fermionic site factors (\ref{eq::statDet}) of heavy-dense QCD in the twisted Potts-model ensembles  (\ref{eq::efe_effPotts}) are mapped to complex external fields, parametrized by  $\chi$, $\eta$, $\eta'$, via
\begin{equation*}
    Q(z) = \e^{\chi + \eta \mathrm{Re}z + \i \eta'\mathrm{Im}z}\,.
\end{equation*}
This mapping is completely analogous to that used in Ref.~\cite{CondellaDetar2000} and for our case explicitly provided in Appendix~\ref{sec::appendix_external_fields}. Some special care must be taken to control numerical errors from the long chains of matrix multiplications for large $L$ in the transfer matrix approach.\footnote{Both, the analytic and the transfer matrix computations of this subsection were done with \emph{Wolfram Mathematica}.$^\text{\textregistered}$}

As an overview we first consider the unrestricted partition function $Z_{\mathrm{eff}}$ and the ensemble $Z(q_V =_3 1)$ with one unit of electric center flux through $S$ for $L=6$ from the analytic calculation. The corresponding total net quark numbers $\langle q \rangle$ 
as functions of the chemical potential $\mu$  are compared in Figure~\ref{fig::plot_fnf_comp}, with $\sigma a/m = 0.3$ and $T/m = 0.1$.  

The maximum site-occupation number of the spin-1/2 fermions in the one-flavor theory with three colors is $2N_c=6$. The unrestricted net quark number from $Z_{\mathrm{eff}}$ (blue) therefore shows the well-known transition of heavy-dense QCD at $\mu=\pm m$, which the tends towards a step-function for $T\ll m$. With increasing positive $\mu$ it changes in one step from $\langle q\rangle = 0$ for the empty lattice, across half filling at $\mu=m$, here with $\langle q\rangle= N_c L =18 $,  to the fully occupied sate at saturation, here with $\langle q\rangle= 2 N_c L=36 $. Neither of these three states contain any fluxes and their energies are thus independent of $\sigma a$. 

 With one unit of electric center flux forced through $S$ in $Z(q_V =_3 1)$ (red) this situation changes in two notable ways: The total net quark number zero configuration is favored only for $|\mu | < m/3$. But even there
the lattice cannot be empty because we need at least one elementary flux $l=1$ at either side of $V$ to generate the required flux in $S^\ast$. In this case, the states that  minimize $\Delta = H -\mu q$  are made of exactly one quark-anti-quark pair where the quark sits inside $V$, either at site $i=0$ or at site $L/2-1$, and is connected to an anti-quark at the adjacent site in $\overline V$ (at $i=L-1$ or $L/2$), cf.~Figure \ref{fig::plot_qnd_1d}~(a).

At $\mu= \pm m/3$ baryonic or anti-baryonic plateaus occur which extend over the respective $\mu$ intervals $(m/3, m)$ or $(-m,-m/3)$. 
The baryonic states that minimize $\Delta$ consist of one quark in $V$ and a diquark at the adjacent site in $\overline{V}$. For $T\ll \sigma a$, the two are again connected by exactly one elementary flux, and the resulting baryon is therefore localized at either side of the boundary between  $V$ and  $\overline{V}$.
The energy difference between the mesonic ($\mathrm{q}\mathrm{\overline{q}}$, favored for $|\mu|< m/3$) and the baryonic states ($\mathrm{qqq}$, for $m/3 < \mu <m $) is just $\Delta(\mathrm{qqq}) - \Delta(\mathrm{q}\mathrm{\overline{q}}) = m-3\mu$ and vanishes at the transition. For $ m/3 < \mu < m$ the baryonic state therefore minimizes $\Delta$ and a new plateau forms. For $\mu > m $ quark-anti-quark pairs fill up the lattice until one baryonic hole is left to accommodate the required flux. Saturation can therefore not quite be reached  with $q_V =_3 1$, but the maximum occupation number here occurs at $\langle q\rangle=  2 N_c L - 3=33 $. The analogous arguments apply  for the anti-baryonic states at $\mu<0$. 

To further illustrate these localized states in the electric flux ensemble and their temperature dependence, we consider the corresponding local densities in form of the site-dependent net quark number $\langle q_i \rangle$ on a somewhat larger chain of length $L=20$ (again with $\sigma a/m = 0.3$). The interface $S^*$ consists of the two links that start at sites $i=0$ (backward) and $i=L/2-1=9 $ (forward), respectively.
Density plots of the net quark numbers $\langle q_i \rangle$ over site $i$ and temperature $T/m $ are shown in Figure \ref{fig::plot_qnd_1d} for three representative values of the chemical potential. These are $\mu = 0$ for the mesonic regime in (a),  $\mu/m = 0.5$ for the baryonic ground state in (b), and $\mu/m = 1.5$ for the maximally occupied ground state in (c). All figures indicate the formation of states that get localized close the surface $S$ as the temperature is lowered. 

At $\mu/m = 0$ in Fig.~\ref{fig::plot_qnd_1d}  (a) they consist of one meson at either of the two links in $S^*$. Both possibilities are equally likely and the average quark numbers therefore approach $\langle q_i \rangle = 0.5$ at the inside boundary of $S$ in $V$ (at site $i=0$ or $i=L/2-1=9$) and $\langle q_i \rangle = -0.5$ at the outside in $\overline{V}$  (at $i=L/2=10$ or $i=L-1=19$) in the low temperature limit, where all other site-occupation numbers approach zero. Increasing the temperature to  $T \sim \sigma a$, longer flux strings are possible and the quark-anti-quark pairs can extend further into the bulk before these localized mesonic states dissolve in thermal quark-anti-quark excitations as temperature is further increased. 

For $\mu/m =0.5$ in Fig.~\ref{fig::plot_qnd_1d} (b) we analogously observe that the net quark numbers tend to $\langle q_i \rangle = 0.5$ at the inside boundary of $S$ in $V$ and $\langle q_i \rangle = 1$ at the outside in $\overline{V}$ in the low temperature limit, corresponding to the two possibilities of having a quark inside and a diquark outside with minimal electric-flux energy. As temperature increases, the flux strings spread into the bulk and eventually dissolve in quark excitations so that the density gets distributed homogeneously over the lattice when $T\gg\sigma a$.

The low temperature limit at $\mu/m =1.5$ in Fig.~\ref{fig::plot_qnd_1d}  (c), on the other hand, shows how the lattice fills up on all but the boundary sites with the maximum number of $2N_c=6$ quarks. The sites at the inside boundary of $S$ now have $\langle q_i \rangle = 5$ and those outside  $\langle q_i \rangle = 5.5$ in the ground state. This is because inside $S$ we must now accommodate a diquark hole at the boundary on either side of $V$ so that we can either have $ q_i  = 4 $ or $6$, depending on where the diquark hole is localized (with an average of $\langle q_i \rangle = 5$), and the corresponding quark hole then yields  $q_i = 5 $ or $6$ (with an average of $\langle q_i \rangle = 5.5$) at the outside boundary of $S$.   Together they form the baryonic hole necessary to create the required unit of electric flux through $S$ with the shortest string possible.\footnote{Because anti-quarks are suppressed exponentially by $(m+\mu)/T$, the baryonic hole is energetically favored over the corresponding flux configuration involving an anti-quark in $\overline V$. At fixed electric-flux energy the former costs $3(\mu-m)$ in energy while the latter costs $2(\mu-m)+(\mu+m) = 3\mu -m $ to create in the ground state.}

\begin{figure}
    \centering
    \includegraphics[scale=0.68]{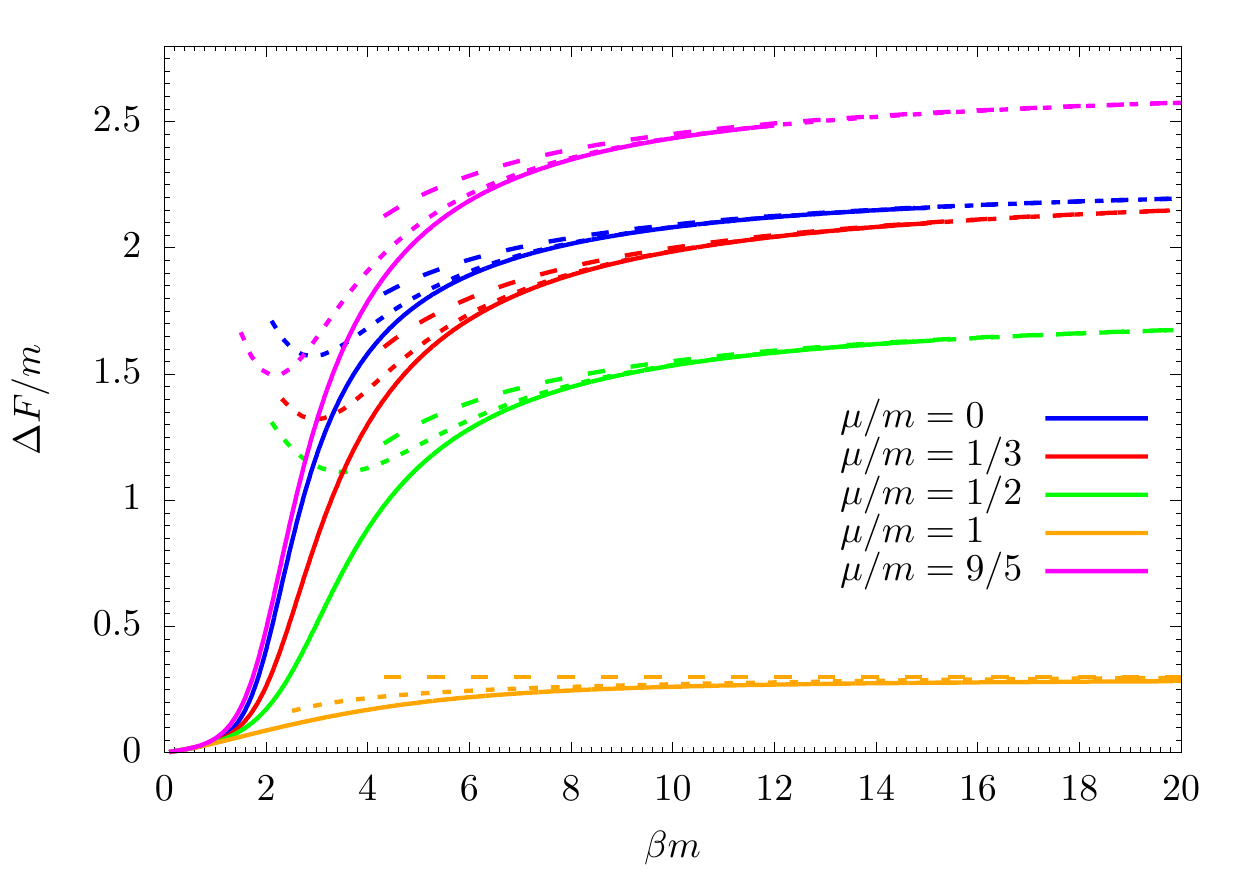}
    \caption{$\Delta F$ of a one-dimensional chain with $L=64$, $\sigma a/m = 0.3$. The solid lines are the results from the transfer matrix approach. The dashed lines represent Eq.~(\ref{eq::approx_free_energy}): without excited state corrections (for $N_1^{(1)}=N_0^{(1)}=0$, long-dashed), and with %contributions from exact
    exact $N_1^{(1)}$, $N_0^{(1)}$ and $\Delta_1^{(1)}$, $\Delta_0^{(1)}$ included (short-dashed).}
    \label{fig::plot_free_energies}
\end{figure}

In order to investigate the electric-flux free energy $\Delta F_\infty$ in (\ref{eq::def_free_energy}), we now use $L=64$ as a proxy for  the $L\to\infty $ limit. 
We have confirmed that this is well justified within the errors for our present purposes, because there are no significant changes between chains with $L\in \{40,44,\ldots,60\}$ and the chain with $L=64$.  The dependence of this $\Delta F$ 
on the inverse temperature $\beta m$ is shown in Figure \ref{fig::plot_free_energies}. As before, the string tension is $\sigma a/m = 0.3$, and the values of the chemical potential are $\mu/m \in\{0,1/3,1/2,1,9/5\}$. The dashed lines represent results from the low-temperature expansion in Equation (\ref{eq::approx_free_energy}), where we have computed the corresponding energy levels  $\Delta_e^{(i)}$ and their multiplicities  $N^{(i)}_e$ exactly. The long-dashed lines show the ground state contributions only, where the leading order $1/\beta $ corrections are determined from the ratio of multiplicities $N^{(0)}_1$ and $N^{(0)}_0$ alone. The short-dashed lines include 
the contributions from the first excited levels with the multiplicities  $N^{(1)}_1$ and $N^{(1)}_0$ to extend this asymptotic expansion.

For the long-dashed leading order we simply have
\begin{equation}
    \Delta_1^{(0)}-\Delta^{(0)}_0 = \begin{cases}
     2m+\sigma a\,, & \mu \le m/3\\
     3 |m-\mu| + \sigma a \, , & \mu > m/3 
    \end{cases} \label{eq::gs_energy}
\end{equation}
for the ground-state energy difference and
\begin{equation}
    \frac{N_1^{(0)}}{N^{(0)}_0} = \begin{cases}
   8 \,, & \mu \le m/3\\
   20\,, & \mu = m/3 \\
   12\,, &  m/3 < \mu \not= m \\
   25/18\,, & \mu = m
    \end{cases} \label{eq::multiplicities}
\end{equation}
with $N_0^{(0)} = 1 $ everywhere but for $\mu=m$ at half filling where $N_0^{(0)} = 6^L $.
The corresponding energies and multiplicities for the corrections from the first excited states are more tedious to work out analytically and not given explicitly, here.  

The different orders of this expansion nicely overlap with one another and with the numerical results from the transfer matrix approach in a reasonably wide range of temperatures to confirm that the exact asymptotic expansions correctly extrapolate the numerical results. We can therefore safely assume the low-temperature limit of $\Delta F_\infty$ to be given by (\ref{eq::low_temp_free_energy}) which tends to  $\Delta_1^{(0)} -\Delta^{(0)}_0$ for $\beta\to\infty $, as expected. In the mesonic regime for $\mu <m/3 $, this difference is simply given by the energy of the quark-anti-quark pair at adjacent sites with one unit of flux in between, cf.~(\ref{eq::gs_energy}).
In the baryonic regime for $\mu>m/3$, it represents the sum of the energy cost $\pm 3m + \sigma a $ of adding or removing a baryon, thereby creating the same flux, and $\mp 3\mu$ for the chemical potential of that baryon or the baryonic hole, depending on whether we are below or above half filling at $\mu=m$. At $\mu=m/3$, the energies for adding the meson or the baryon are exactly the same and the combinatorical factors of both possibilities (8 for the meson and 12 for the baryon) simply add up (hence we have $N_1^{(0)} = 20$ there).

In the high-temperature limit, $\beta \rightarrow 0$, all curves approach zero which is is not quite trivial, as we have seen in the previous subsection, but numerically confirms our discussion there as well.

\subsection{Three-Dimensional Lattice}

As the next step, we investigate $\Delta F_\infty$ in three spatial dimensions, where exact calculations and the direct transfer matrix approach are no-longer feasible. In this section we therefore compute the electric-flux free energies with a  Metropolis algorithm for the Potts-model formulation with complex external fields, where the action of the interface ensembles in Eq.~(\ref{eq::efe_effPotts}) is given by
\begin{equation*}
    \begin{split}
    S_S(k;\{z_i\}) = 2\gamma &\sum_{\langle i,j\rangle}\Re\left(z^{-s_{\langle i,j\rangle} k } \, z_i z_j^\ast\right)\\
    &+\,\eta\sum_{i}\Re z_i + \mathrm{i}\eta'\sum_i \Im z_i \,.
    \end{split}
\end{equation*}
As before, the inverse temperature $\beta=1/T$ always refers to the flux-tube model and is dual to the Potts-model coupling $\gamma$ from Eq.~(\ref{eq::Potts_gamma}).  
The flux tube model parameters $\beta\mu$ and $\beta m$, on the other hand, are again mapped onto $\eta$ and $\eta'$ as described in Appendix \ref{sec::appendix_external_fields}. This mapping implies $\eta' = 0$, if $\mu = 0$. 
When $\mu \neq 0$, the action is complex, in general. To avoid any more or less severe sign problem that this might cause, here we simply restrict to the case of vanishing chemical potential, $\mu = 0$. The free energy is computed from the ratio of $Z(q_V =_3 1)$ and $Z(q_V =_3 0)$ which we rewrite as ratios of observables relative to the untwisted ensemble $Z_S(0)=Z_\mathrm{eff}$. This reweighting yields
\begin{align}
    \Delta F &=-\frac{1}{\beta}\ln\left(\frac{1-\frac{1}{2}\left(\langle \e^{S_1 - S_0}\rangle_0+\langle\e^{S_2 - S_0}\rangle_0\right)}{1+\langle \e^{S_1 - S_0}\rangle_0+\langle \e^{S_2 - S_0}\rangle_0}\right)\,, \label{eq::rewight}
\end{align}
with the shorthand notation $S_k\equiv S_S(k;\{z_i\})$. The expectation values $\langle \bullet \rangle_0$ are taken over the untwisted ensemble $Z_S(0)=Z_\mathrm{eff}$, and we have used the fact that $Z(q_V =_3 e)$ and $Z_S(k)$ are both real when $\mu = 0$ and hence $S_k \in \mathbb{R}$.

\begin{figure}
    \centering
    \includegraphics[scale=0.68]{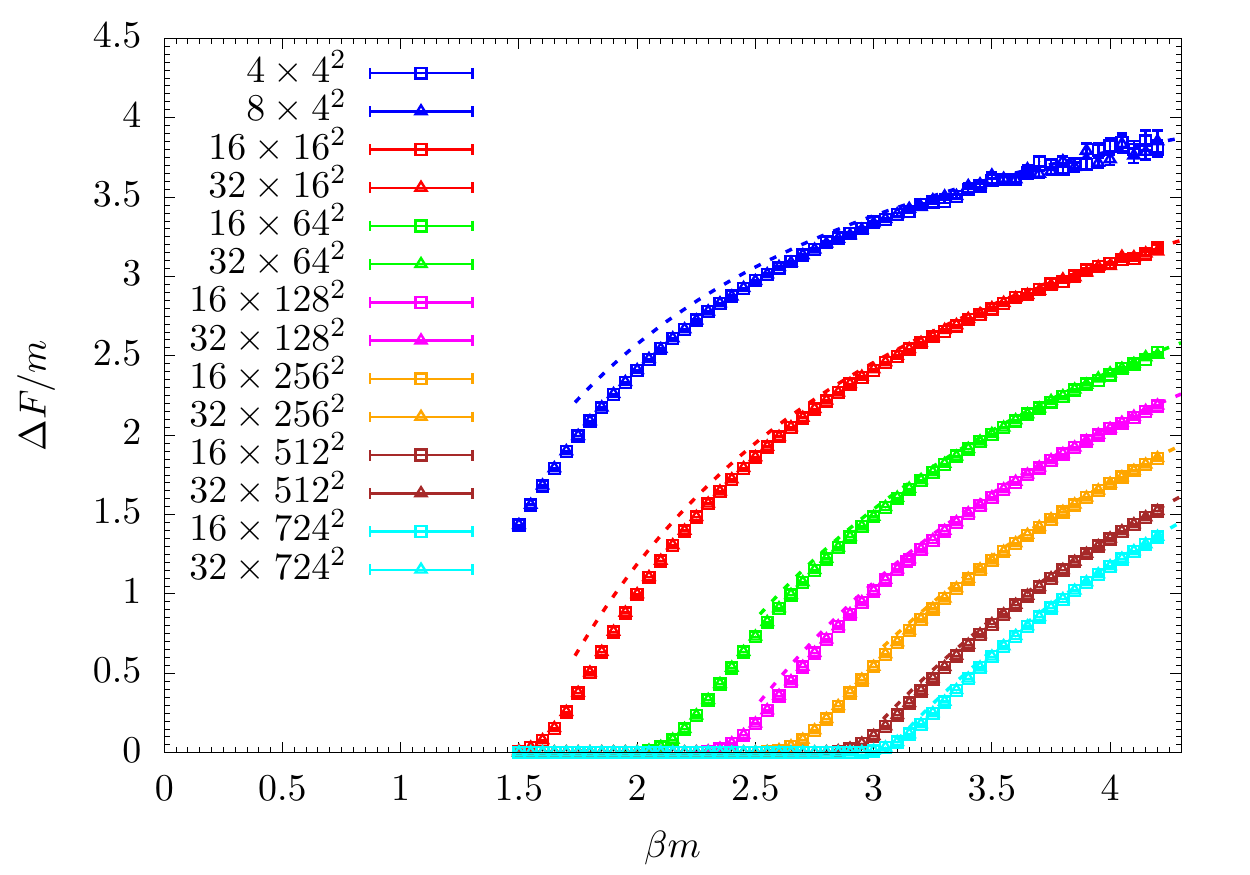}
    \caption{$\Delta F$ for three-dimensional lattices of various sizes and $\sigma a/m = 3$, dashed lines represent Eq.~(\ref{eq::approx_free_energy}) with $N_1^{(1)}=N_0^{(1)}=0$, lattice sizes are labelled by $L_1\times (L_2L_3)$.}
    \label{fig::free_energies_pre_3}
\end{figure}

\begin{figure}
    \centering
    \includegraphics[scale=0.68]{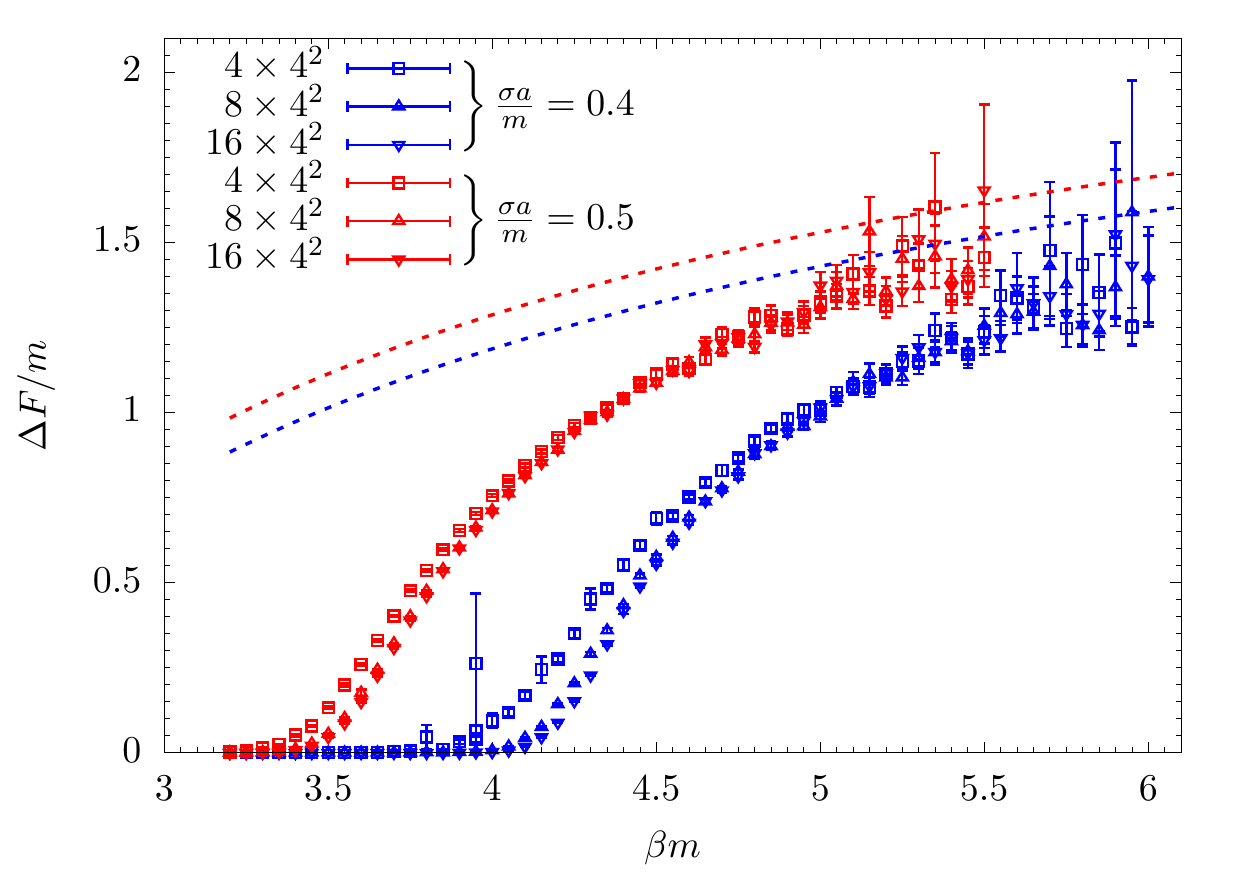}
    \caption{
    $\Delta F$ for three-dimensional lattices of various sizes with dashed lines from Eq.~(\ref{eq::approx_free_energy}) with $N_1^{(1)}=N_0^{(1)}=0$ as in
    labels as in Fig.~\ref{fig::free_energies_pre_3}, but here for 
     $\sigma a/m = 0.5$ (red) and $\sigma a/m = 0.4$ (blue).}
    \label{fig::free_energies_combined}
\end{figure}

We have computed the electric-flux free energies $\Delta F$ for $\sigma a/m = 3$,  $\sigma a/m =  0.5$ and $\sigma a/m = 0.4$. The results are summarized in Figures \ref{fig::free_energies_pre_3} and \ref{fig::free_energies_combined}. 
With the strong string tension, $\sigma a = 3 m$, the results in Figure \ref{fig::free_energies_pre_3} are practically independent of $L_1$, already for the rather small lattices: results with different $L_1$ are plotted with different symbols (hardly distinguishable in Fig.~\ref{fig::free_energies_pre_3}) but in the same color for the same values of $L_2$ and $L_3$. In all cases the free-energy differences of the electric fluxes vanish for $\beta\to 0$  in the high-temperature limit, as in the one-dimensional case.
The new feature here is that the results depend on the transverse surface area $A_\perp = L_2L_3$ of each of the two interfaces. In fact, at any constant but non-zero temperature, the electric-flux free energies tend to zero with $A_\perp \to \infty$. This is due to the entropy that arises from the increasing number of possibilities to place a localized mesonic state at the surface $S=\partial V$ (with its flux in $S^*$).

At any finite $A_\perp$, however, the zero-temperature limit remains finite as given by
the asymptotic behavior in Equation (\ref{eq::approx_free_energy}) and shown with dashed lines in Fig.~\ref{fig::free_energies_pre_3} (for the ground state contributions, setting  $N_1^{(1)}=N_0^{(1)}=0$). We can hence conclude that 
\begin{equation}
    \label{eq::low_temp_approx_three_dim}
    \Delta F \simeq (\Delta_1^{(0)} - \Delta_0^{(0)}) -\frac{1}{\beta}\ln\left(8 A_\perp \right)\, ,
\end{equation}
with the zero temperature result $\Delta_1^{(0)} -\Delta^{(0)}_0$ as in one dimension, and independent of $A_\perp=L_2L_3 $ in this order of limits. And as in (\ref{eq::multiplicities}) before, the factor of $8$ arises because there are four possibilities to arrange a localized quark-anti-quark pair
with one unit of flux, such that the quark is in $V$ and the anti-quark in $\overline{V}$, at some place in the two interfaces of size $A_\perp$ that make up $S^{\ast}$ (the area of $S$ is $2A_\perp $).

This implies that both ensembles are nearly equivalent when the area of the interface   gets sufficiently large. The presence of the additional mesonic quark-anti-quark structure, which creates the required flux through $S^\ast$, becomes insignificant relative to the background configurations of the $Z(q_V =_3 0)$ ensemble. 
In fact, under the assumption that the required flux  through $S^\ast$ is created from localized  structures, the low-temperature behavior of the free energy $\Delta F$ has an intuitive interpretation: If we assign a penetration depth $d$ of this structure perpendicular to $S$  into the bulk, as determined from the string-breaking scale, we expect that the $L_1$ dependence vanishes, if $L_1 \gg d$. The two interfaces of $S^*$ decorrelate. With the large $\sigma a=3 m$ this is basically the case for all values of $L_1$ in Figure \ref{fig::free_energies_pre_3}. In order to explicitly see a dependence on $L_1$ we therefore show analogous results with smaller values of $\sigma a = 0.5 m$ (red) and $\sigma a = 0.4 m$ (blue) in Figure \ref{fig::free_energies_combined}, where the approach to the limit $L_1\to\infty$ can be seen in the crossover regions. The downside of the smaller values of $\sigma a$  is that the signal is lost earlier, with the simple reweighting in (\ref{eq::rewight}), due to an overlap problem at lower temperatures where $\Delta F$ grows larger. There is reasonable evidence, however, that the noisy data of Figure \ref{fig::free_energies_combined} nevertheless approaches the analytic limit of Eq.~(\ref{eq::low_temp_approx_three_dim}), again indicated by dashed lines.

\begin{figure}
    \centering
    \includegraphics[scale=0.68]{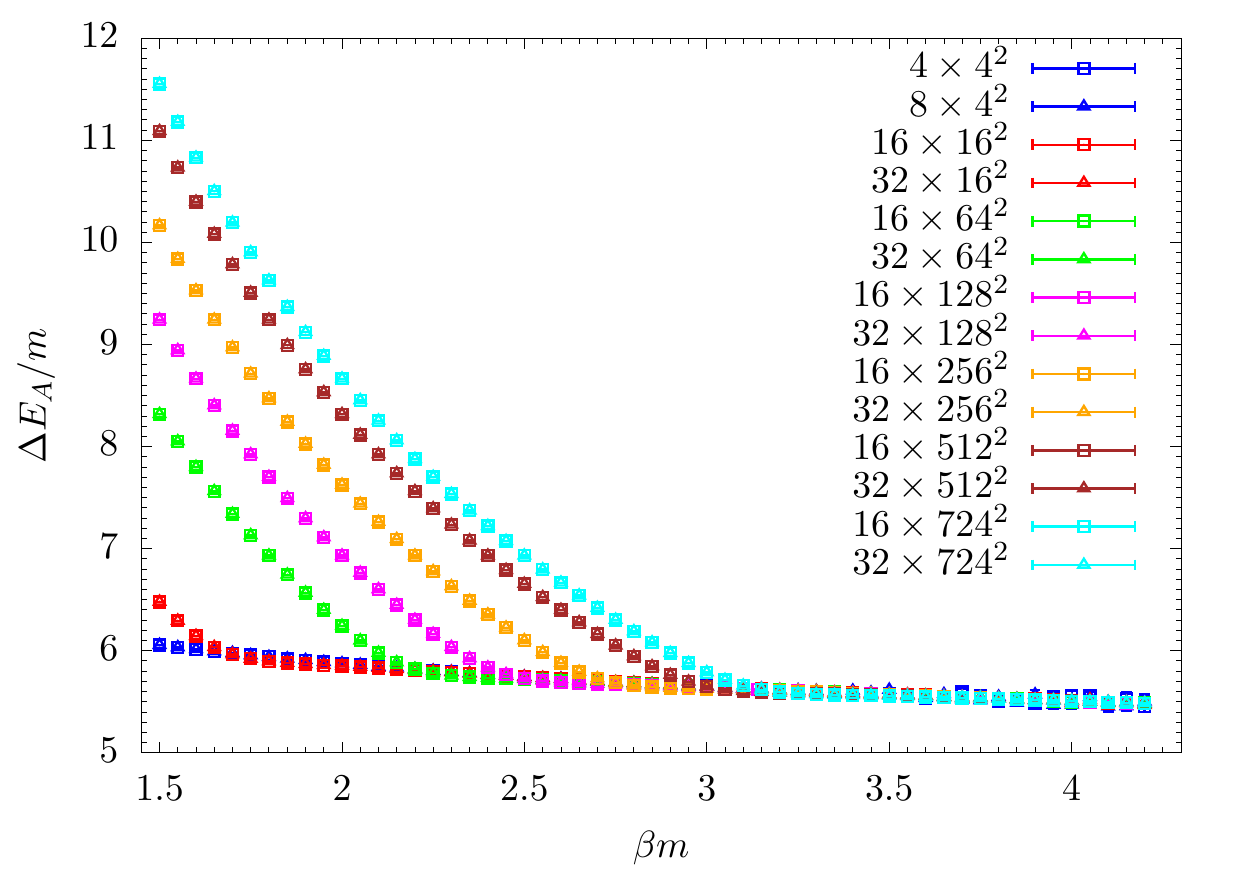}
    \caption{$\Delta E_{A}$ in (\ref{eq::DEA} for the results of Figure \ref{fig::free_energies_pre_3}. Lattice sizes labelled by $L_1\times (L_2L_3)$. With $\sigma a =3m$ here, the zero-temperature limit is thus given by $2m+\sigma a = 5 m$.} 
    \label{fig::energies_a}
\end{figure}

 To summarize our conclusions from Figures \ref{fig::free_energies_pre_3} and \ref{fig::free_energies_combined}, we approximate the free energy difference, $\Delta F = \Delta E - T\Delta S$, by Equation (\ref{eq::low_temp_approx_three_dim}) in the low temperature limit. Hence, we can identify the leading entropy difference as $\Delta S = \ln\left(8 A_\perp \right)$ with $A_\perp = L_2L_3$ in this limit. 
 We therefore define 
 \begin{equation}
    \Delta E_A = \Delta F + \frac{1}{\beta}\ln(8A_\perp)\, , \label{eq::DEA}
\end{equation} 
to test whether this effective energy approaches a constant value, independent of $A_\perp$, at sufficiently small temperatures. 
The result  for $\sigma a  = 3 m$ is shown in Figure \ref{fig::energies_a}. 

The data for all different $A_\perp$ indeed collapse at sufficiently low temperatures, and the values for $\Delta E_A$ become independent of $A_\perp$ within precision.
The temperature for this collapse to occur decreases with increasing $A_\perp$.
Note that for arbitrarily large areas our assignment of entropy and energy must eventually break down to avoid negative free energy. At lower temperature this breakdown occurs for larger interfaces. The most natural explanation is that more complicated flux configurations with more complicated combinatorical factors
 contribute and our simple assignment of entropy from the possible translations of the elementary quark-anti-quark pair across the interface breaks down.

\section{Towards Lattice QCD}
\label{sec::towards_lattice_qcd}

In this section we describe the generalization of our construction of ensembles with quark numbers $q_V\not= 0\!\mod 3 $ inside a finite spatial volume $V$ from our effective  flux-tube model for heavy-dense QCD to full Lattice QCD with dynamical quarks. We will proceed in two steps. First, we undo the midpoint definition for the group integrations of the Polyakov loops with the reduced Haar measure in (\ref{eq::midpdef}) and reintroduce the spatial link variables again, in order to demonstrate how the construction immediately carries over to heavy-dense QCD in the next subsection. 
When spatial hops are possible, i.e.~beyond the static fermion determinant of the leading order in the underlying hopping expansion, another subtlety arises because one then has to restrict the dynamics in such a way 
that the net-quark number modulo three in the volume $V$ remains unchanged at all times. Intuitively, this will be achieved by introducing a selectively permeable static  membrane which will allow only hadrons to pass back and forth but not individual quarks or diquarks. Luckily, however, this will then be the last conceptual step that is necessary for the generalization to full QCD. In Subsection~\ref{subsec::full_qcd}  we approach this problem with the dualization scheme described in \cite{GattringerMarchis2017,MarchisGattringer2018} which directly provides the suitable generalizations of the notions of fluxes and local net quark numbers in the flux-tube model, based on loops of dual variables winding around the Euclidean time direction. 

Here, we consider Wilson fermions instead of staggered fermions and only employ the dualization of the fermionic part. The derivations are otherwise completely analogous to \cite{GattringerMarchis2017,MarchisGattringer2018} and, in particular, so are the graphs to represent the dual-variable configurations by closed loops. 

The general setup starts from the partition function which is formally given by
\begin{equation*}
    Z = \int\mathcal{D}U\mathcal{D}\overline{\psi}\mathcal{D}\psi\,\e^{S_G(U)+S_F(\overline{\psi},\psi, U)} \,,
\end{equation*}
with integrals over the gauge-link variables $U_{\langle x,y\rangle} \equiv U_{x,\mu} $ from site $x$ to $y=x+\hat \mu $ (where $x$, $y$ are the spacetime lattice indices, we will continue to reserve $i$, $j$ for spatial lattice indices),
and on-site Grassmann generators $\overline{\psi}_{x}^{\omega}$ and $\psi_{x}^{\omega}$ with $\omega $ combining color and Dirac indices. 
For  our purposes it will be sufficient to use the standard Wilson plaquette action as our gauge action $S_G$,
\begin{equation}
    \label{eq::plaquette_action}
    S_G = \frac{2}{g_0^2}\sum_{x,\mu < \nu} \ReTr U_{\partial p_{x,\mu\nu}}\,,
\end{equation}
with the product of link variables along the boundary of an oriented plaquette $p_{x,\mu\nu}$ which is given by
\begin{equation*}
    \begin{split}
    \partial p_{x,\mu\nu} = \{\langle x, x+ &\hat{\mu}\rangle,\langle x +\hat{\mu}, x+ \hat{\mu}+\hat{\nu}\rangle,\\
    &\langle  x+ \hat{\mu}+\hat{\nu}, x+\hat{\nu}\rangle,\langle x+\hat{\nu},x\rangle\}\,.
    \end{split}
\end{equation*}
For chemical potential in the Wilson fermion action $S_F$, we use the standard Hasenfratz-Karsch prescription \cite{HasenfratzKarsch1983},
\begin{equation}
    \label{eq::fermion_action}
    S_F = \sum_{x, \mu}\kappa\bigg[\overline{\psi}_x\Lambda^{+}_{ x,\mu}\psi_{x+\hat{\mu}}+\overline{\psi}_{x+\hat{\mu}}\Lambda^{-}_{x,\mu}\psi_{x}\bigg]-\sum_x \overline{\psi}_x\psi_x\, ,
\end{equation}
with
\begin{align}
    \Lambda^{+}_{x, \mu} &= ((1-\delta_{\mu,4}) + \e^{+ a\mu}\delta_{\mu,4})(1 -  \gamma_\mu)
    %U_{\langle i, i+\hat{\mu}\rangle}\,,\\
    U_{x,\mu}\,,  \nonumber \\
    \Lambda^{-}_{x,\mu} &= ((1-\delta_{\mu,4}) + \e^{-a\mu}\delta_{\mu,4})(1+\gamma_\mu)
    %U_{\langle i+\hat{\mu},i\rangle}\,,
    U_{x,\mu}^\dagger\,, \nonumber
\end{align}
hermitian Euclidean gamma matrices, $\{\gamma_\mu,\gamma_\nu\} = 2 \delta_{\mu\nu} $, and the hopping parameter $\kappa$. The gauge fields are periodic in all directions, and the Grassmann variables are anti-periodic in the Euclidean time direction (and periodic in spatial directions), as usual.

\subsection{Heavy-Dense Limit}

In Section \ref{sec::electrix_flux_ensembles}, we have seen how concepts analogous to  those used in the construction of  't~Hooft's electric-flux ensembles can be used in the flux-tube model to fix the electric flux (mod $3$) through a closed surface $S =\partial V$, by twisting the Potts-model couplings along the directed links through $S$, which then fixes the total net quark number (mod $3$) in the volume $V$ inside $S$. The first step in this sequence is to replace center-vortex sheets in the gauge theory by interfaces forcing cyclic shifts in the Potts model which then, after $\mathrm{Z}_3$-Fourier transform, generate the flux through $S$ required for the non-trivial charge inside, from the $\mathbb{Z}_3$-Gauss law in the dual and hence equivalent flux-tube model. 

\begin{figure}
    \centering
    \includegraphics[scale=0.45]{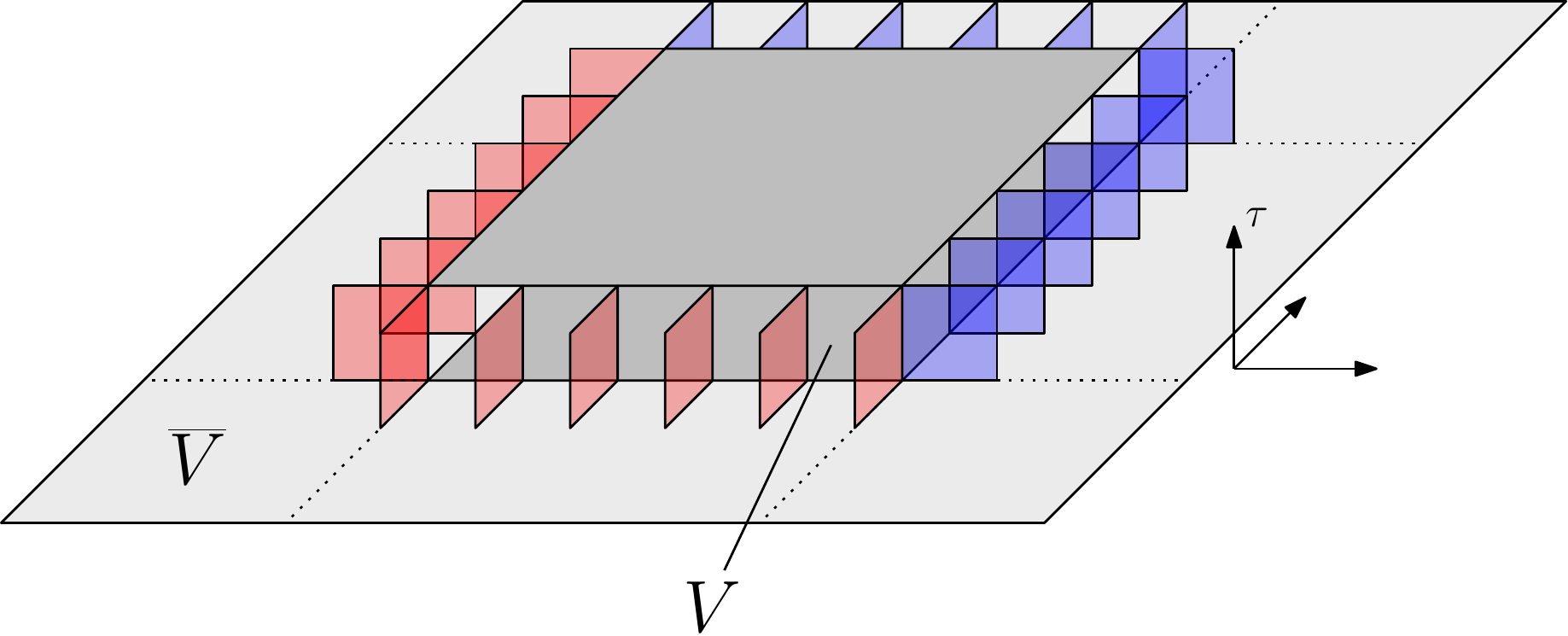}
    \caption{Two Dirac sheets of opposite twist enclosing the spatial volume $V$ in one layer of Euclidean time. The sheets represent the twists of the action $S_G^z$. Blue plaquettes are twisted with $z$ and red plaquettes are twisted with $z^{-1}$.}
    \label{fig::twisted_su3_static}
\end{figure}

Therefore, by analogy, we might first simply introduce the closed center-vortex sheet in the plaquette action.
In the gauge theory this is done by twisting a stack of temporal plaquettes with spatial edges normal to $S = \partial V$. This can be realized by two Dirac sheets of opposite twists enclosing the volume $V$ between one pair of adjacent time slices as is illustrated in Figure~\ref{fig::twisted_su3_static} for the  $(2+1)$-dimensional case.
The ensemble of fixed quark number would then be defined as
\begin{equation}
    \label{eq::fixed_partition_naiv}
    Z(q_V =_3 e) = \frac{1}{3}\sum_{z\in\mathrm{Z}_3}z^{-e}\int\mathcal{D}U\mathcal{D}\overline{\psi}\mathcal{D}\psi\,\e^{S_G^z + S_F}
\end{equation}
with the twisted plaquette action denoted by $S_G^z$.

In the pure gauge theory, when there is no fermion action in the first place, the invariance of the integration measure can be used to remove the closed center-vortex sheets so that  $S_G^z \to S_G $ in each term of this sum, and the final $\mathrm{Z}_3$-Fourier transform produces an exact  zero whenever $q_V\not= 0\!\!\mod 3$.
 
This already changes, however, once we include the static fermion determinant of heavy-dense QCD. The transformation that removes the closed center-vortex sheets from $S_G^z$,
which multiples all temporal links between the two time slices in $V$ by a center element, at the same time now also changes the site factors of the 
static fermion determinant in $V$,  
\begin{equation}
    \det(Q_{\mathrm{stat}}) \to \det(Q_{\mathrm{stat}}^z) = \left[\,\prod_{i\in V}Q(zL_i)\right]\left[\,\prod_{i\notin V}Q( L_i)\right]\,. \nonumber
\end{equation}
Closed center-vortex sheets can no-longer be removed anymore, and the Ansatz~(\ref{eq::fixed_partition_naiv})
therefore indeed provides the desired generalization of the Potts/flux-tube model construction  of the previous section, to fix $q_V \not= 0 \!\!\mod 3$.
Intuitively, at this leading order in the underlying hopping expansion it is sufficient to fix the net quark number in $V$ at any one instant of time because it can never change  without spatial hops in the static fermion determinant.

The corresponding effective Polyakov-loop theory follows  from standard arguments \cite{LangelageLottini2011,LangelageNeuman2014}, cf.~Sec.~\ref{sec::eft-ft}.
When the gauge action is approximated at the lowest order of the strong-coupling expansion by the nearest-neighbour Polyakov-loop interaction, the corresponding $Z_3$-Fourier transform over the closed center-vortex sheet becomes
\begin{equation}
    Z_{\mathrm{eff}}(q_V =_3 e) = \frac{1}{3}\sum_{z\in\mathrm{Z}_3}z^{-e}\int\mathcal{D}L\, \e^{S_{\mathrm{eff}}^z}\det(Q_{\mathrm{stat}})
    \label{eq::strong-hop-q-part}
\end{equation}
with
\begin{equation*}
    S_{\mathrm{eff}}^z =\sum_{\langle i,j \rangle}2\gamma\,\Re\left(z^{-s_{\langle i,j\rangle}}L_i L^\ast_j \right)\,.
\end{equation*}
In particular, approximating the gauge-group integrations by the sum of the midpoints of the three center sectors in the reduced Haar measure, we arrive precisely at Eqs.~(\ref{eq::efe_quark_fourier}), (\ref{eq::efe_effPotts}) again (with $z^k \to z \in \mathrm Z_3$ here).

While the construction in  (\ref{eq::strong-hop-q-part}) is not restricted to the strong-coupling expansion but remains valid with the full plaquette action included, we reiterate, however, that it is sufficient only for static fermions. In full Lattice QCD the closed center-vortex sheets must be introduced between all time slices to prevent the net quark number in the volume $V$ from fluctuating, as we will show next.

\subsection{Full Lattice QCD}
\label{subsec::full_qcd}

The exponential of the fermion action can be decomposed into a product of exponentials of $\overline{\psi}_x^{\omega}\psi_y^{\omega'}$. The expansion of each such exponential stops at order one. The possible powers  $k_{xy}^{\omega\omega'}\in\{0,1\}$ define the dual variables. Expanding all resulting factors, and after integration of the gauge-link variables, the partition function can formally be written as a sum over configurations of the dual variables as follows:
\begin{equation}
    \label{eq::dual_partition_qcd}
    \begin{split}
        Z &=\int\mathcal{D}[\ldots]\,\e^{a\mu \sum_x(n_x - \overline{n}_x)}\Bigg[\prod_{x,\omega}\left(-\overline{\psi}_x^{\omega}\psi_x^{\omega}\right)^{k_{xx}^{\omega\omega}}\Bigg]\\
        &\hskip .2cm  \times\Bigg[\prod_{x,\mu,\omega,\omega'}\left(\overline{\psi}_x^\omega\psi^{\omega'}_{x+\hat{\mu}}\right)^{k_{x,x+\hat{\mu}}^{\omega\omega'}}\left(\overline{\psi}_{x+\hat{\mu}}^\omega\psi^{\omega'}_{x}\right)^{k_{x+\hat{\mu},x}^{\omega\omega'}}\Bigg] \, ,
    \end{split}
\end{equation}
with
\begin{equation}
    \label{eq::euclidean_particle_number}
    n_x = \sum_{\omega,\omega'}k_{x,x+\hat{4}}^{\omega\omega'}\,,\quad\overline{n}_x = \sum_{\omega,\omega'}k_{x+\hat{4},x}^{\omega\omega'} \, ,
\end{equation}
and the short-hand notation
\begin{equation*}
    \int\mathcal{D}[\ldots] =  \sum_{\{k\}}W(\{k\})\int\mathcal{D}\overline{\psi}\mathcal{D}\psi\,.
\end{equation*}
One then observes that only dual on-site and nearest-neighbor variables occur in the partition function. The weights $W(\{k\})$ formally result from integrating the gauge links for each given configuration of dual variables and include the Boltzmann factor with the plaquette action. 

The Grassmann integrations are non-zero if and only if the monomials in Equation~(\ref{eq::dual_partition_qcd}) contain each generator exactly once. This constrains the configurations $\{k\}$ that can be realized, and leads to the following rules \cite{GattringerMarchis2017,MarchisGattringer2018} for diagrammatically representing the monomials that contribute to the partition function in Equation~(\ref{eq::dual_partition_qcd}): 
Products  $\overline{\psi}_x^\omega\psi_y^{\omega'}$ of nearest-neighbor pairs
are represented by arrows starting at site $x$ and ending at $y= x \pm \hat\mu$. The direction of the arrow therefore indicates whether 
 $k_{x,x+\hat{\mu}}^{\omega\omega'} = 1$ (forward arrow) or $k_{x+\hat{\mu},x}^{\omega\omega'} = 1$ (backward arrow). 
A pair $(\overline{\psi}_x^\omega, \psi_x^{\omega})$ at the same site $x$ occuring in a monomial is represented by one open circle at site $x$ for each $\omega $. The arrows therefore start and end in these circles by construction. When $k_{xx}^{\omega\omega} = 1$ for some $\omega $, representing the presence of an on-site pair $-\overline{\psi}_x^\omega \psi_x^\omega$ in the monomial, then this circle is blocked for arrows and hence represented as a filled one. Because at every site, every generator has to occur exactly once, every circle in a vertex is thus either occupied and hence filled, or exactly one arrow has to start and another one has to end in the otherwise open circle. It is then obvious that the dual variables have to form closed loops. A closed loop either consist of distinct arrows (connecting open circles) or a filled circle, which can also be considered as a loop of zero length. 

The number of arrows ending at a site $x$ must equal the number of arrows starting there. Hence, we get the following conservation law:
\begin{equation}
    \sum_{\mu,\omega,\omega'}\left(k_{x-\hat{\mu},x}^{\omega\omega'} - k_{x,x-\hat{\mu}}^{\omega\omega'}+k_{x+\hat{\mu},x}^{\omega\omega'}-k_{x,x+\hat{\mu}}^{\omega\omega'}\right) = 0\,. \label{eq::qcons}
\end{equation}
This represents the continuity equation for the net quark number on the lattice. To see this explicitly, we use the definition of the local quark and anti-quark numbers in Equation~(\ref{eq::euclidean_particle_number}) to define the net quark number $q_x = n_x -\bar n_x $, and introduce the corresponding quark-number current,
\begin{equation*}
j_{x,x+\hat k} =  \sum_{\omega\omega'}\left(k_{x,x+\hat k}^{\omega\omega'}-k_{x+\hat k ,x }^{\omega\omega'}\right)\, ,
\end{equation*}
in the spatial direction $\hat k$. Then, splitting the sum over $\hat\mu $ in (\ref{eq::qcons}) into $\hat\mu = \{\hat k,\hat 4\}$ yields  
\begin{equation}
    \label{eq::continuity_equation}
    \Delta_4 q_x  = -\sum_{k=1}^3\left(j_{x,x+\hat{k}}-j_{x-\hat{k},x}\right)
\end{equation}
where $ \Delta_4 q_x = q_x - q_{x-\hat{4}} $ is the change of the local net quark number in Euclidean time, and the right hand side the discrete spatial divergence of the associated quark-number current $\vec j$ at $x$. If we also split the site index $x=(i,\tau)$ into spatial $i$ and temporal coordinate index $\tau$, we can rewrite 
Equation (\ref{eq::continuity_equation}) to show that the total net quark number 
$q_\tau = \sum_i(n_{(i,\tau)}-\overline{n}_{(i,\tau)}) $
is conserved with periodic spatial boundary conditions,
\begin{equation*}
    \Delta_4 q_\tau = \sum_i\Delta_4 q_{(i,\tau)} = 0 \, .
\end{equation*}
Hence, we can also write
\begin{equation*}
    \e^{\beta\mu q_\tau} = \e^{a\mu\sum_x(n_x-\overline{n}_x)}
\end{equation*}
because the total net quark number is independent of the Euclidean time $\tau$. In particular, this fugacity relation with the chemical potential $\mu$ thus confirms the correct identification of the local (anti-)quark numbers with the dual flux variables in the temporal direction in Equation~(\ref{eq::euclidean_particle_number}).
 
Any loop of dual variables with timelike extend can thus be interpreted as quarks and anti-quarks propagating in positive time direction, because its timelike links increase the values of the corresponding occupation numbers $n_x$ and $\overline{n}_x$. Obviously only loops of dual variables winding around the Euclidean time direction can contribute to $q_\tau$ which agrees with the interpretation of net quark number as winding loops in \cite{MarchisGattringer2018}.  
%Such a loop can be interpreted as a quark or anti-quark starting at some spatial site $\vec{x}$ and hopping along the spatial links only to end at $\vec{x}$ once the maximal time is reached.

\begin{figure}
    \centering
    \includegraphics[scale=1.0]{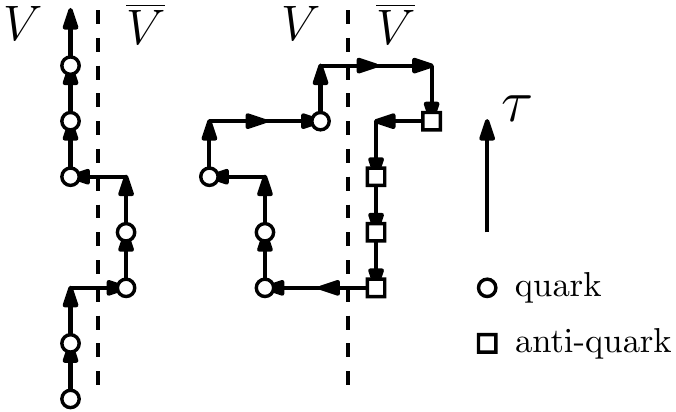}
    \caption{Fermion loops with timelike contributions interpreted as quarks and anti-quarks propagating in the positive direction of Euclidean time: individual quarks can hop back and forth between   $V$ and  $\overline{V}$ (left), or hadronic clusters can be created and annihilated across the boundary between the two (right).}
    \label{fig::fermion_paths}
\end{figure}

Next, we consider a spatial subvolume $V$ with boundary $S = {\partial} V$ as in the flux-tube model, and the net number $q_{V,\tau} $ of quarks in $V$ at the Euclidean time $\tau $. For its change in Euclidean time we have Gauss's law in the form,
\begin{equation}
    \Delta_4 q_{V,\tau} = \sum_{i\in V}\Delta_4 q_{(i,\tau)} = -\sum_{\mathclap{\langle x,x+\hat k \rangle \in (S^\ast,\tau)}}\,j_{x,x+\hat k} \equiv -\Phi_{S,\tau}\,, \label{eq:anotherGauss}
\end{equation}
where $(S^\ast,\tau)$ denotes the set of spatial links that intersect $S$ to connect $V$ with its complement $\overline V $ in time slice $\tau$. 

In contrast to the constant total net quark number $q_\tau$, the values of $q_{V,\tau}$ can change in time because the total flux $\Phi_{S,\tau}$ in (\ref{eq:anotherGauss}), of net charge through $S$  (which is not the same as the center flux $\phi_S$ from the previous section), in not necessarily zero, which just means that loops of dual fluxes can leave or enter $V$. This corresponds to the hopping of quarks in and out of $V$, or the simultaneous creation or annihilation of quarks and anti-quarks on either side of $S$, as illustrated in Figure~\ref{fig::fermion_paths}.  To fix the quark number  $q_{V,\tau}$
(mod $3$) once and for all, it can therefore not be enough to introduce an interface in form of a closed center-vortex sheet between only one pair of subsequent time slices, as it was in the heavy-dense limit.
As we will see explicitly from the transfer-matrix formulation in the next section, we have to constrain the dynamics such that the subspace of our state-space in which the net quark number in $V$ is fixed remains unchanged as we evolve from one time slice to another. 
To achieve this here, we have to fix the net quark number in each time slice by suitably constraining the sum over the dual variables, defining
\begin{equation}
    \begin{split}
        Z(q_V=_3 e) =\int\mathcal{D}[\ldots]\,&\e^{a\mu q_\tau}\Bigg[\prod_{\tau}\delta_3(q_{V,\tau} - e)\Bigg]\\
        \times&\Bigg[\prod_{x,\omega}\ldots\Bigg]\Bigg[\prod_{x,\mu,\omega,\omega'}\ldots\Bigg]\,.
    \end{split} \label{eq::cZedual}
\end{equation}
Using the definition of the Kronecker delta modulo three, $\delta_3(\phi)$ in (\ref{eq::Kdm3}), this becomes
\begin{equation*}
    Z(q_V=_3 e) =\frac{1}{3^{L_4}}\sum_{z_0\ldots,z_{L_4-1}}\Bigg[\prod_{\tau}z_\tau^{-e}\Bigg]\, Z(\{z_\tau\}) \, %\label{eq::ZqV}
\end{equation*}
where the twisted partition functions are given by
\begin{equation*}
    \begin{split}
    Z(\{z_\tau\}) = \int\mathcal{D}[\ldots]&\Bigg[\prod_{\tau}(\e^{a\mu})^{ q_{\overline{V},\tau}}(\e^{a\mu}z_\tau)^{q_{V,\tau}}\Bigg]\\
    \times&\Bigg[\prod_{x,\omega}\ldots\Bigg]\Bigg[\prod_{x,\mu,\omega,\omega'}\ldots\Bigg]\,.
    \end{split}
\end{equation*}
Note that only the local net quark numbers $q_{(i,\tau )}$ with $i \in V $ get multiplied by the center elements $z_\tau \in \mathrm Z_3$, $\tau = 0,\dots L_4-1$, which is equivalent to  adding 
local imaginary parts  $2\pi k_\tau/3 $ with the corresponding $k_\tau \in \mathbb Z_3 $ to the chemical potential in $V$.   
Undoing the steps for the dualization, these center elements therefore multiply the corresponding temporal link variables in the fermion action,  
\begin{equation*}
    Z(\{z_\tau\}) = \int\mathcal{D}U\mathcal{D}\overline{\psi}\mathcal{D}\psi\,\e^{S_G(U)+S_F(\{z\},U,\overline{\psi},\psi)}
\end{equation*}
with the modified fermion action
\begin{equation*}
    \begin{split}
        S_F&(\{z_\tau\},U,\overline{\psi},\psi) = \\
        &\sum_{x, \mu}\kappa\bigg[\overline{\psi}_x\Gamma^{+}_{ x,\mu}\psi_{x+\hat{\mu}}+\overline{\psi}_{x+\hat{\mu}}\Gamma^{-}_{x,\mu}\psi_{x}\bigg]-\sum_x \overline{\psi}_x\psi_x
    \end{split}
\end{equation*}
where we have introduced
\begin{align*}
    \Gamma_{x,\mu}^{\pm} = \begin{cases}
        z_\tau^{\pm 1} \Lambda^{\pm}_{(i,\tau),4}\,, & i \in V\text{ and }\mu = 4\\
         \Lambda^{\pm}_{x,4}\,, &\text{otherwise}
    \end{cases}\,,
\end{align*}
i.e.~the timelike forward/backward hopping terms between time slices $\tau$ and $\tau +1$ get multiplied with center elements $z_\tau$/$z_\tau^{-1}$, if the corresponding temporal link lies in $V$.
Formally, for these links this amounts to replacing $\e^{\pm a \mu} \to z_\tau^{\pm 1} \e^{\pm a \mu} $ in the fermion action.

\begin{figure}
    \centering
    \includegraphics{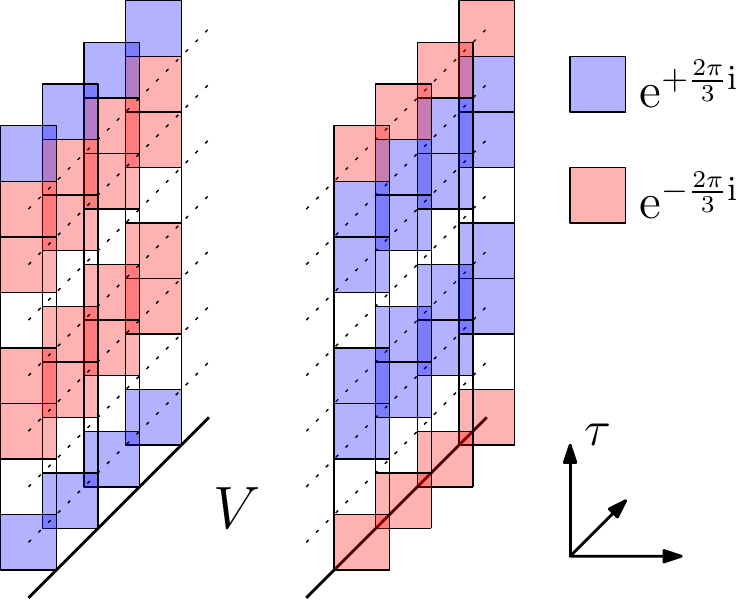}
    \caption{Exemplary illustration of a twist configuration in the plaquette action $S_G(\{z_\tau\}, U)$ in $2+1$ dimensions with $\tau$ indicating  the direction of Euclidean time (white plaquettes represent cases with $z_\tau = 1$).}
    \label{fig::twisted_su3}
\end{figure}

We can now use the invariance of the Haar measure to substitute  $U_{x,4} \to z_\tau^{-1} U_{x,4} $ for the temporal links at $x=(i,\tau)$ with $i\in V$, to transfer the set of center elements $\{z_\tau\}$ here from the fermion action into the gauge action, in the reverse manner already used for heavy-dense QCD in the previous subsection. 
This restores the standard Hasenfratz-Karsch form of the fermion action again, %The transformation is given by $U_{\langle i, i+\hat{4}\rangle} \rightarrow z_\tau^{-1} U_{\langle i, i+\hat{4}\rangle}$ for all $i = (\vec{x},\tau)$ with $\vec{x}\in V$. We arrive at
and we arrive at
\begin{equation}
    Z(\{z_\tau\}) = \int\mathcal{D}U\mathcal{D}\overline{\psi}\mathcal{D}\psi\,\e^{S_G(\{z_\tau\}, U) + S_F(U,\overline{\psi},\psi)}\,, \label{eq::ZqVpS}
\end{equation}
where we now have the twisted plaquette action
\begin{equation*}
    S_G(\{z_\tau\},U) = \frac{2}{g_0^2}\sum_{x,\mu<\nu} \ReTr(z(p_{x,\mu\nu})U_{\partial p_{x,\mu\nu}})\,,
\end{equation*}
with
\begin{equation}
    z(p_{(i,\tau),\mu\nu}) = \begin{cases}
        \hskip 6pt z_\tau, & \nu = 4, \mu = k, \langle i, i +\hat{k}\rangle \in S^\ast \\
        z_\tau^{-1}, & \nu = 4, \mu = k, \langle i +\hat{k}, i \rangle \in S^\ast\\
        \hskip 11pt 1, & \text{otherwise}
    \end{cases}\,. \label{eq::tpS}
\end{equation}
As compared to the heavy-dense limit in the previous subsection, the important extension here is that the temporal plaquettes with spatial edges in $S^*$ get twisted between \emph{all} time slices as illustrated for $2+1$ dimensions in Figure~\ref{fig::twisted_su3}.

If we consider only static quarks again, i.e.~ignore spatial hopping,  then  the total flux of net charge through $S$ always vanishes, we have  $\Phi_{S,\tau} = 0$ and hence $\Delta_4 q_{V,\tau} = 0$ in Eq.~(\ref{eq:anotherGauss}). In this case, it is therefore sufficient to fix the quark number $q_V$ in $V$ in any one layer between Euclidean time slices. The other constraints are then fulfilled automatically, and the general construction of this subsection then simply boils down to Equation~(\ref{eq::fixed_partition_naiv}) used for the static fermion action of the previous subsection.

In summary, to fix the net quark number in a finite volume $V$ to some value $q_V= e$ (mod $3$), we have to restrict the dynamics such that only multiples of three quarks and anti-quarks can enter and leave the volume at any time. To achieve this, it is not enough to insert twists in only a single layer between subsequent time slices, as in the construction of 't~Hooft's electric flux ensembles in the pure gauge theory. The necessary generalization is surprisingly simple, however. We just have to twist the temporal plaquettes with spatial edge in $S^*$ (i.e.~dual to $S=\partial V$ in three dimensions) between \emph{all} time slices in the gauge action. In particular, the fermion determinant remains unchanged, and standard Hybrid-Monte-Carlo simulation techniques in combination with the snake algorithm developed for the 't~Hooft loops in the pure gauge theory \cite{ForcrandDElia2001}, although expensive, should be possible at least in principle. The $\mathrm Z_3$-Fourier transforms in all time slices of the twisted plaquette actions illustrated in Figure~\ref{fig::twisted_su3}, over all possible $\{z_\tau\} $ configurations (corresponding to $3^{L_4}$ different combinations of closed center vortex sheets), then introduce the selectively permeable static membrane to block quarks but not hadrons from fluctuating in and out of $V$ as described in the introduction to this section.

\section{Transfer-Matrix Construction}
\label{sec::trans_const}

The transfer-matrix formulation of Lattice QCD  \cite{BorgsSeiler1983, Luescher1977,Palumbo2002,Mitrjushkin2002,Mitrjushkin2003,Smit2002} is based on the maximal temporal gauge, in which all temporal lattice links are unity except for those between one single layer of subsequent Euclidean time slices which determines the Polyakov loops. 
A Hilbert space $\mathcal{H}$ is then defined as a tensor product of  square integrable functions $ L^2(\mathrm{SU}(3))$ on $\mathrm{SU}(3)$ for every spatial link variable $U_{\langle i, i+\hat k\rangle}$ on the lattice, with the Fock space generated by fermionic creation and annihilation operators,
\[ (\hat{\xi}_{\mathrm{q}}^\sigma)^\dagger_{i,a}, \; (\hat{\xi}_{\mathrm{q}}^\sigma)_{i,a} \;\; \mbox{and}\;\;  (\hat{\xi}_{\overline{\mathrm{q}}}^\sigma)^\dagger_{i,a}, \;  (\hat{\xi}_{\overline{\mathrm{q}}}^\sigma)_{i,a}\,  \]   
for  quarks and anti-quarks with spin $\sigma= \{\uparrow,\downarrow\} $ and color $a$ on spatial lattice site $i$.
Combining their different labels in multi-indices, a state $\ket{\psi}\in\mathcal{H}$ can be decomposed into
\begin{equation*}
    \ket{\psi} = \sum_n\sum_{i_1,\ldots,i_n}f_{i_1,\ldots,i_n}(U)\, \hat{\xi}^\dagger_{i_1}\ldots\hat{\xi}^\dagger_{i_n}\ket{0} \, ,
\end{equation*}
with complex-valued functions $f_{i_1,\ldots,i_n}(U)$ on the set of spatial gauge configurations $U = \{U_{\langle i,i+\hat{k}\rangle}\}$.

The residual gauge invariance consists of time-indepen\-dent transformations  
$\hat{\varrho}(\Omega)$, with $\Omega_i \in\mathrm{SU}(3)$ which act on tensor fields $f(U)$ and Grassmann generators as follows, 
\begin{align*}
    (\hat{\varrho}(\Omega)f)(U) &= f(\{\ldots,\Omega^\dagger_i U_{\langle i, i+\hat{k}\rangle}\Omega_{i+\hat{k}},\ldots\})\,,\\
    \hat{\varrho}(\Omega)(\hat{\xi}_{\mathrm{q}}^\sigma)_{i,a}\hat{\varrho}^\dagger(\Omega) &= (\Omega^\dagger_i)^{ab}(\hat{\xi}_{\mathrm{q}}^\sigma)_{i,b}\,,\\
    \hat{\varrho}(\Omega)(\hat{\xi}_{\overline{\mathrm{q}}}^\sigma)_{i,a}\hat{\varrho}^\dagger(\Omega) &= (\hat{\xi}_{\overline{\mathrm{q}}}^\sigma)_{i,b}(\Omega_i)^{ba}\,,
\end{align*}
where contracted indices are summed. We denote with $\mathcal{H}_{\mathrm{phys}}\subseteq \mathcal{H}$ the subspace of all gauge-invariant and hence physical states: $\hat{\varrho}(\Omega)\ket{\psi} = \ket{\psi}$ for all gauge transformations $\Omega$ and  $\ket\psi \in \mathcal{H}_{\mathrm{phys}}$.
An intuitive example of a physical state is a quark-anti-quark pair connected by a line of gauge fields \cite{KogutSusskind1975}:
\begin{equation*}
    \ket{\psi} = (\hat{\xi}^\sigma_{\mathrm{q}})^\dagger_i\left[U_{\langle i,i_1\rangle}\ldots U_{\langle i_{n},j\rangle}\right](\hat{\xi}^\gamma_{\overline{\mathrm{q}}})^\dagger_j \, \ket{0}\,.
\end{equation*}
To project onto the subspace of gauge-invariant and physical states, one defines a projection operator $\hat{P}_0$ via
\begin{equation}
    \label{eq::proj_gauge_inv}
    \hat{P}_0\ket{\psi} = \int\mathcal{D}\Omega\,\hat{\varrho}(\Omega)\ket{\psi}\,.
\end{equation}
When the transfer operator $\hat{T}$ maps physical states onto physical states, it commutes with $\hat{P}_0$, and 
\begin{equation}
    \label{eq::partition_trans_const}
    Z = \tr\left(\e^{\beta\mu\hat{N}}\hat{T}^{L_4}\hat{P}_0\right)
\end{equation}
defines the Lattice QCD partition function with chemical potential $\mu$, total net quark number operator $\hat{N}$, and of extend $L_4$ in the discrete Euclidean time direction. 

The transfer operator $\hat{T}$ is commonly expressed in terms of a gauge-field integral with kernel $K(U,U')$ deduced from its action on products of gluonic and fermionic  states represented by $f(U)$ and $\ket{\psi_F}$, %in $\mathcal H$, %(\ref{eq::fock-space-exp}),
\begin{equation*}
    \hat{T}\big(f(U)\otimes\ket{\psi_F}\big) = \int\mathcal{D}U'\,K(U,U')\big(f(U')\otimes\ket{\psi_F}\big) \, .
\end{equation*}
In Luescher's original formulation  \cite{Luescher1977}, the  kernel $K(U,U')$ is split symmetrically into gluonic and fermionic factors.
For our later purposes here, on the other hand,  it will be more convenient to factor  $\hat{T}$ in an asymmetric way of the 
following form,
\begin{equation}
    \begin{split}
        K(U,&U') = S(U,U')T_G(U')T_F(U')\,, 
    \end{split}\label{eq::askern}
\end{equation}
where $U$ and $U'$ are gauge field configurations again, the gauge factors $S$, $T_G$  and the fermionic $T_F$  are given explicitly in Appendix~\ref{sec::appendix_trans_const} for completeness.

Although manifestly different from that in  \cite{Luescher1977}
this integral kernel nevertheless yields the same partition function, and the resulting transfer operators are hence thermodynamically equivalent, as shown in Appendix~\ref{sec::appendix_trans_const}.
While also still gauge invariant, our asymmetric transfer operator is unfortunately not the exponential of a hermitian Hamiltonian, however. It is hence not necessarily strictly positive anymore, but its eigenvalues can be complex in general. As a weaker assumption we must therefore argue that this asymmetric transfer operator does lead to a hermitian Hamiltonian $\hat{H}$ at least in the continuous-time limit, so that introducing the Euclidean-time discretization $a_4= \beta/L_4 $, we can write
\begin{equation}
    \label{eq::continuum_trans_op}
    \big(\hat{T}(a_4) \big)^{L_4} \longrightarrow \e^{-\beta\hat{H}}\; , \;\; L_4\rightarrow\infty \,.
\end{equation}
In Appendix \ref{sec::euclidean_time_continuum} we show that the linear order in $a_4$ of the transfer operator $\hat{T}(a_4)$
yields a gauge-invariant and hermitian Hamiltonian $\hat{H} = \hat{H}_{\mathrm{G}}+\hat{H}_{\mathrm{F}}$ where $\hat{H}_{\mathrm{G}}$ is the purge gauge part and $\hat{H}_{\mathrm{F}}$ the fermionic part which turns out to be given by the standard Hamiltonian for Wilson fermions  on the lattice (\cite{MontvayMuenster1994}). This can therefore 
be seen as a heuristic justification of Equation (\ref{eq::continuum_trans_op}). Non-hermitian transfer operators have previously been studied for improved gauge actions  \cite{LuescherWeisz1984}, for example, and  asymmetric transfer operators have also been used in Ref.~\cite{AokiIizuka2017}.

The next step is to introduce the concept of $\mathbb{Z}_3$-charges and $\mathbb{Z}_3$-electric fluxes. The latter are defined on link $l$ , as in Ref.~\cite{BorgsSeiler1983}, through
\begin{equation*}
    (\hat{E}^z_{l}f)(U) = f(Z_l U) \, ,
\end{equation*}
where    
\begin{equation*}
Z_{l\langle i,j \rangle}  =    \begin{cases}
        z^\dagger, & \langle i,j\rangle = l\\
        \hskip 4pt z, & \langle j,i \rangle = l \\
        \hskip 4pt 1, & \mbox{otherwise} 
    \end{cases} \, , \;\; \mbox{and} \;\; z\in \mathrm{Z}_3\,.
\end{equation*}
The charge operators are defined on site $i$ by 
\begin{align*}
    \hat{Q}_i^z (\hat{\xi}_{\mathrm{q}}^\sigma)^\dagger_{i,a}(\hat{Q}^z_i)^\dagger  &= z\cdot (\hat{\xi}_{\mathrm{q}}^\sigma)^\dagger_{i,a}\,,\\
    \hat{Q}_i^z (\hat{\xi}_{\overline{\mathrm{q}}}^\sigma)^\dagger_{i,a}(\hat{Q}^z_i)^\dagger &= z^\dagger\cdot (\hat{\xi}_{\overline{\mathrm{q}}}^\sigma)^\dagger_{i,a}\,,
\end{align*}
that is, quarks are sources of positive center charge and anti-quarks are sources of negative center charge. We then consider states $\ket{\psi}\in \mathcal{H}$ with  well-defined values of fluxes and charges: We say $\ket{\psi}$ has a center-electric flux $e\in\mathbb{Z}_3$ in direction of the link $l$ if and only if
\begin{equation}
    \label{eq::state_center_field}
    \hat{E}_{l}^z \ket{\psi} = z^e \ket{\psi}
\end{equation}
for all $z\in\mathrm{Z}_3$. Analogously, the state has a center charge $q\in\mathbb{Z}_3$ at the site $i$ if and only if
\begin{equation}
    \label{eq::state_center_charge}
    \hat{Q}^z_i\ket{\psi} = z^q\ket{\psi}
\end{equation}
for all $z\in\mathrm{Z}_3$. A center operator is a unitary representation of $\mathrm{Z}_3$, and states with flux $e$ or charge $q$ span irreducible representations of $\mathrm{Z}_3$. Conversely, all irreducible representations of a center operator are one-dimensional by Schur's lemma and have to act as in Eqs.~(\ref{eq::state_center_field}) and (\ref{eq::state_center_charge}). Hence, for a center operator, all one-dimensional subspaces with a center value are exactly the irreducible representations of the operator. The subspace $\mathcal{H}_{\mathrm{phys}}$ of physical configurations is invariant under the action of all center operators because they commute with $\hat{P}_0$ as can be seen from its definition in (\ref{eq::proj_gauge_inv}). We therefore restrict all center operators to $\mathcal{H}_{\mathrm{phys}}$ in everything discussed below.

The concept of center fields and charges can now be used to establish a local $\mathbb{Z}_3$-Gauss law, and to classify the Hilbert space in terms of sectors of center charge in the volume $V$. This construction is based on essentially the same ideas as in \cite{Mack1978, KijowskiRudolph2002, KijowskiRudolph2002-2, KijowskiRudolph2005}. Here, we establish the charged sectors by decomposing the Hilbert space into subspaces associated with flux-tube configurations $\{q,e\} $. Employing the Peter-Weyl theorem we write
\begin{equation}
    \mathcal{H}_{\mathrm{phys}} = \bigoplus_{\{q,e\}}\mathcal{H}_{\{q,e\}}\,, \label{eq::decompose1}
\end{equation}
where $\mathcal{H}_{\{q,e\}}$ denotes the subspace of all states which carry a configuration $\{q,e\}$ of center values, i.e.~such states simultaneously provide irreducible representations of all charge and field operators. For the interpretation of these states as flux-tube states, we also need a local $\mathbb{Z}_3$-Gauss law as follows: If $\ket{\psi}\in\mathcal{H}_{\mathrm{phys}}$, it is possible to write
\begin{equation}
    \label{eq::gauss_transfer_matrix}
    \hat{Q}_i^z\prod_{j\sim i} \hat{E}^z_{\langle i, j \rangle}\ket{\psi} = \ket{\psi}\, ,
\end{equation}
for all $z\in\mathrm{Z}_3$ because the product of the center operators constitutes a local gauge transformation. This is precisely our local $\mathbb{Z}_3$-Gauss law again, however, which in this form generalizes that derived for the pure gauge theory in Ref.~\cite{BorgsSeiler1983}. 
For all $\ket{\psi}\in \mathcal{H}_{\{q,e\}}$, on the other hand, Equation~(\ref{eq::gauss_transfer_matrix}) implies
\begin{equation*}
    z^{q_i + \sum_{j\sim i} e_{\langle i, j\rangle }} \ket{\psi} = \ket{\psi}\,, \;\;  z\in\mathrm{Z}_3 \, ,
\end{equation*}
which for $\ket{\psi} \not= 0$ implies
\begin{equation}
    q_i + \sum_{j\sim i}  e_{\langle i, j\rangle } = 0 \! \mod 3\, .
    \label{eq::Z3-GaussQe}
\end{equation}
Hence, one either has $\mathcal{H}_{\{q,e\}} = \{0\}$ or the local $\mathbb{Z}_3$-Gauss law must be satisfied at all sites of $\{q,e\}$. We can therefore safely restrict the decomposition in (\ref{eq::decompose1}) to all center configurations $\{q,e\}_\mathrm{phys} $ that obey the local $\mathbb{Z}_3$-Gauss law (\ref{eq::Z3-GaussQe}) without loss,
\begin{equation*}
    \mathcal{H}_{\mathrm{phys}} = \bigoplus_{\{q,e\}_{\mathrm{phys}}}\mathcal{H}_{\{q,e\}}\,.
\end{equation*}
As a result, we now have a decomposition of all physical states into sums of states which each are associated with some physical flux-tube configuration $\{q,e\}_\mathrm{phys}$. In particular, this classification of flux-tube states, in the same way as in our flux-tube model, also allows to decompose the Hilbert space into sectors of fixed center charge in a subvolume $V$:
\begin{equation*}
    \mathcal{H}_{\mathrm{phys}} = \mathcal{H}_{q_V =_3\,0} \oplus \mathcal{H}_{q_V =_3\,1} \oplus \mathcal{H}_{q_V =_3\,2}
\end{equation*}
where
\begin{equation*}
    q_V = \sum_{i \in V} q_i \,.
\end{equation*}
The projection operator onto a subspace $\mathcal{H}_{q_V =_3 e}$ is
\begin{equation*}
    \hat{P}_{q_V}(e)  = \frac{1}{3}\sum_{z\in\mathrm{Z}_3}z^{-e}\prod_{i \in V}\hat{Q}_i^z\,.
\end{equation*}
Inserting the $\mathbb Z_3$-Gauss law in (\ref{eq::gauss_transfer_matrix}), we obtain
\begin{align}
    \hat{P}_{q_V}(e) =& \frac{1}{3}\sum_{z\in\mathrm{Z}_3}z^{-e}\prod_{ i \in V}\prod_{j\sim i} \hat{E}_{\langle i,j\rangle}^{z^{-1}}\nonumber\\
    =& \frac{1}{3}\sum_{z\in\mathrm{Z}_3}z^{e}\prod_{\langle i,j\rangle\in S^*}\hat{E}_{\langle i,j\rangle}^{z} \label{eq::chargefluxproj}\\
    =& \frac{1}{3}\sum_{z\in\mathrm{Z}_3}z^{e}    \, 
    \hat\phi_S^z \equiv  \hat{P}_{\phi_S}(-e)\,, \nonumber
\end{align}
where we have introduced the operator $\hat\phi_S^z$ for the total flux through $S=\partial V$ in the last line.
Therefore, we immediately obtain that $\mathcal{H}_{q_V =_3 e} = \mathcal{H}_{\phi_S =_3 -e}$, i.e.~the center charge in the volume determines the flux through its boundary and vice versa, as required.

The alert reader will have noticed that we have tacitly used the same notation for the center charges in (\ref{eq::state_center_charge}) as for the net quark numbers in the previous section. To understand that this was well justified, consider the subspace $\mathcal H_{N_V=_3 e} $ where the total net quark number $N_V$ modulo three in $V$ is fixed, i.e.~the closure of the linear span of gauge-invariant eigenstates of $\hat{N}_{V}$ with net quark number $N_V  = e\,\mathrm{\mod}\,3$. It is an easy exercise to show that the two are indeed the same,  $\mathcal{H}_{N_V =_3 e} = \mathcal{H}_{q_V =_3 e}$, i.e.~the sectors of fixed center charges correspond to the sectors of fixed quark numbers (mod $3$).

Now that we have successfully constructed the flux projection operators $\hat{P}_{\phi_S}(e)$ in the last line of (\ref{eq::chargefluxproj}),  we note that the transfer operator $\hat T$ does not in general commute with those, $\mathcal{H}_{q_V =_3 e}$ is not invariant under $\hat{T}$. This means that if we take an initial state $\ket{\psi} \in\mathcal{H}_{q_V =_3 e}$, contributions from the other charge sectors will be generated by the dynamics.
For the intuition, consider $\ket{\psi}$ to be a simultaneous eigenstate of all quark number operators $\hat{N}_{\mathrm{q},i}$ and anti-quark number operators $\hat{N}_{\overline{\mathrm{q}},i}$. The application of $\hat{T}$ then in general produces non-vanishing contributions $\ket{\psi_{q_V =_3 s}}$ from other charged sectors $s\not= e$. Such contributions are in the orthogonal complement of $\mathcal H_{q_V=_3 e} $ of states whose net quark numbers modulo three must be reshuffled between $V$ and $\overline{V}$ as compared to those in the initial state $\ket\psi $. As in the previous section, we can think of different kinds of changes in the occupation numbers $n_{\mathrm{q},i}$ and $n_{\overline{\mathrm{q}},i}$. For example, individual quarks, anti-quarks or (anti-)diquarks might have moved between $V$ and $\overline{V}$, or an extended quark-anti-quark/diquark hadronic cluster might have been produced across the interface separating  $V$ and $\overline{V}$. The transfer operator includes such processes, because the dynamics are blind to our subvolume $V$. For 't Hooft's electric-flux ensembles in the pure gauge theory, this problem is not present because there are no fermions and the analogous projection operator for the sectors of fixed electric flux therefore commutes with the transfer operator \cite{BorgsSeiler1983}. 
In full QCD the situation is more complicated: While the center-charge sectors correctly define subspaces of fixed net quark numbers (mod $3$), the restriction to those sectors is not automatically compatible with the dynamics. We are therefore forced to restrict the dynamics across the surface $S$ such that the subspace is invariant under the correspondingly modified transfer operator $\hat{T}'$ which must also be gauge-invariant, of course. Moreover, the partition function
\begin{equation}
    Z(q_V =_3 e) = \mathrm{tr}\left(\e^{\beta\mu\hat{N}}(\hat{T}')^{L_4}\hat{P}_{\phi_S}(-e)\hat{P}_0\right) \label{eq:tforpf}
\end{equation}
should equal the one derived in the previous section. To achieve this, we here propose the following modified transfer operator
\begin{equation}
    \label{eq::modified_trans_op}
    \hat{T}' = \frac{1}{3}\sum_{z\in\mathrm{Z}_3} \hat{\phi}_S^z \hat{T}\hat{\phi}_S^{z^{-1}}\,.
\end{equation}
First, the operator $\hat{T}'$ is gauge-invariant because the center-flux operators $\hat\phi_S^z$ commute with $\hat{\varrho}(\Omega)$. Secondly, for $\ket{\psi}\in\mathcal{H}_{\phi_S =_3 -e}$, we also have
\begin{equation}
    \label{eq::inv_charged_sector}
    \ket{\psi'}=\hat{T}'\ket{\psi} = \hat{P}_{\phi_S}(-e)\hat{T}\ket{\psi}\in\mathcal{H}_{\phi_S =_3 -e}
\end{equation}
which establishes the invariance of $\mathcal{H}_{\phi_S=_3 -e}$ under $\hat{T}'$. 

Note that the arguably simpler definition $\hat{T}' = \hat{P}_{\phi_S}(-e)\hat{T}$ will also leave  $\mathcal{H}_{\phi_S =_3 -e}$ invariant, but not the other center-charge sectors. Such a definition would thus depend on the sector which is not good enough. 
We want to modify the dynamics at the surface $S$ such that all sectors are left invariant under the same dynamics. The invariance of the center-charge sectors should be an intrinsic property of the dynamics rather than it being our choice which sector is left invariant. This important property is certainly realized in the definition of Eq.~(\ref{eq::modified_trans_op}). Moreover,  just as our asymmetric transfer operator $\hat{T}$ in Eq.~(\ref{eq::continuum_trans_op}), also $\hat{T}'$ has a well-defined Euclidean time continuum limit, as shown in Appendix~\ref{sec::euclidean_time_continuum}, with the gauge-invariant and hermitian Hamiltonian 
\begin{equation*}
    \hat{H}' = \frac{1}{3}\sum_{z\in\mathrm{Z}_3}\hat{\phi}_S^z \hat{H}\hat{\phi}_S^{z^{-1}}
\end{equation*}
under which the center-charge sectors remain invariant. Therefore, the choice of $\hat{T}'$ is again well justified from the quantum mechanical point of view, describing a quantum mechanical system in which the dynamics of $\hat{H}$ is modified locally at the surface $S$.

Finally, it only remains to be shown that the partition function in (\ref{eq:tforpf}) is equivalent to that of the previous section. Equation~(\ref{eq::inv_charged_sector}) first implies,
\begin{equation*}
    Z(q_V =_3 e) = \mathrm{tr}\left(\e^{\beta\mu\hat{N}}\hat{P}_0(\hat{P}_{\phi_S}(-e)\hat{T})^{L_4}\right)\, .
\end{equation*}
With the definition of the flux projection operators in (\ref{eq::chargefluxproj}) (here with substituting $z_\tau \to z_\tau^{-1} $ in the $Z_3 $ sums for later convenience)
%(and $\hat{P}_{\phi_S^z = -e} = \hat{P}_{\phi_S^{z^{-1}} = e} $)  
this then yields
\begin{align*}
        Z(q_V =_3 e) 
        = \frac{1}{3^{L_4}}\sum_{\{z_\tau \in Z_3\}}&\Bigg[\prod_{\tau=0}^{L_4-1} z_\tau^{-e}\Bigg]\\
        \times &\, \mathrm{tr}\left(\e^{\beta\mu\hat{N}}\hat{P}_0\prod_{\tau=0}^{L_4-1}\left[\hat{\phi}_S^{z^{-1}_\tau}\hat{T}\right]\right)\, ,
\end{align*}
 where the products for $\tau = 0$ to $\tau = L_4-1$ inside the trace are ordered from left to right. The operator $\phi_S^{z^{-1}}\hat{T}$ is obtained from the integral kernel $K(U^z,U')$, with
\begin{equation*}
    U^z_{\langle i,j\rangle} = z^{s_{\langle i,j\rangle}} U_{\langle i,j \rangle} \,.
\end{equation*}
Now our asymmetric definition of the kernel in Eq.~(\ref{eq::askern}) finally becomes handy, because the non-trivial center elements then only multiply links in the first gauge factor.
In $S(U^z,U')$ on the other hand, they simply result in precisely the same twisted plaquette gauge action as in the previous section, when the maximal temporal gauge is relaxed in the end again, after the trace of each term has been evaluated just as in the untwisted case, cf.~Appendix~\ref{sec::appendix_trans_const}. In total, by the standard sequence of steps from the transfer-matrix formulation to the path-integral representation, one then quite obviously arrives at the same result as in the previous section, 
\begin{align}
        Z(q_V=_3 e) =& \frac{1}{3^{L_4}}\sum_{\{z_\tau\in Z_3\}}\Bigg[\prod_{\tau}z_\tau^{-e}\Bigg]\, \label{eq::ZqVefinal}\\
        &\times \int\mathcal{D}U\mathcal{D}\overline{\psi}\mathcal{D}\psi\,\e^{S_G(\{z_\tau\}, U) + S_F(U,\overline{\psi},\psi)}\,, \nonumber 
\end{align}
with the standard Wilson-fermion action in (\ref{eq::fermion_action}), and the twisted plaquette action  
given by Eqs.~(\ref{eq::ZqVpS}) -- (\ref{eq::tpS}).

It is therefore possible to define a gauge-invariant transfer operator $\hat{T}'$ under which the subspaces $\mathcal{H}_{q_V =_3 e}$ with $e\in \mathbb Z_3$  remain invariant. The ensembles over $\mathcal{H}_{q_V =_3 e}$ are equivalent to those defined by constraining the dual variables from  (\ref{eq::cZedual}) in the previous section. Via Eq.~(\ref{eq::modified_trans_op}) the transfer operator  $\hat{T}'$ is obtained from the unconstrained $\hat{T}$ by the unitary operators   $\hat{\phi}_S^z$ for the center-electric flux through $S=\partial V$ which only modify the spatial links in the coboundary $S^\ast$. This local change in the dynamics at the interface $S^\ast$ between the subvolume $V$ and its complement  $\overline V$ here amounts to introducing the same selectively permeable static membrane, which prevents quarks but not hadrons from fluctuating in and out of $V$, as that obtained from constraining the dual variables in the previous section. Practically, to introduce such an interface, one needs to introduce closed center-vortex sheets  around the spatial subvolume $V$ between \emph{all} subsequent time slices. This thus is a more severe modification than that needed for the electric-flux ensembles in the pure gauge theory, but at least it can be realized with only changing the gauge action, owing to our asymmetric transfer operator $\hat T$, from the asymmetric kernel defined in (\ref{eq::askern}). The validity of the latter, on the other hand,  is justified by the fact that $\hat{T}'$ has a well-defined Euclidean-time continuum limit with an Hamiltonian $\hat{H}'$ obtained from the corresponding unmodified $\hat{H}$ via the same operators $\hat\phi_S^z$  for center-flux through the surface $S$. 

Finally we emphasize that no claim is made here as to the uniqueness of our construction. For example, one might choose the symmetric integral kernel of the original Ref.~\cite{Luescher1977} and hence define a different $\hat{T}'$ for the constrained dynamics via Equation (\ref{eq::modified_trans_op}) starting from that. 
In this case, one would have a non-trivial influence of the projection operators $\hat{P}_{\phi_S}(-e)$ on the fermionic part. The path integral would end up looking very different from 
(\ref{eq::ZqVefinal}), including very non-trivial modifications to the fermionic action. The equivalence of such an approach to the asymmetric formulation used here, and the dualization of the previous section is far from obvious. It can hold only in the Euclidean-time continuum limit in the first place, in which both transfer operators yield the same Hamiltonian $\hat{H}$ and therefore, from the construction in  (\ref{eq::modified_trans_op}),  the same modified  $\hat{H}'$, likewise. Guided by the results from the dualization of the fermion determinant, we have thus proposed but one specific form for the  modified transfer operator $\hat{T}'$, with the local modifications interpreted in terms of conjugation with the center-flux operators $\hat{\phi}_S^z$.

\section{Conclusion}
\label{sec::conclu}

We have constructed ensembles $Z(q_V =_3 e)$ with fixed net quark numbers $q_V =  e\! \mod 3$ (fractional baryon number) in a spatial subvolume $V$ for full Lattice QCD (here with one flavor of Wilson fermions). As shown explicitly in Sections \ref{sec::towards_lattice_qcd} and \ref{sec::trans_const}, our construction is predominantly based on two major concepts: (a) the existence of a local $\mathbb{Z}_3$-Gauss law in QCD, and (b), the possibility of modifying the dynamics at the surface $S = \partial V$ of the spatial subvolume $V$. The local $\mathbb{Z}_3$-Gauss law is thereby crucial to establish the equivalence between the center-electric flux through $S$ and the net quark number modulo three in $V$, which is fixed by enforcing the corresponding center flux through $S$. At this surface $S$, however, this then requires modifying the dynamics of QCD  which would otherwise allow 
individual quarks, anti-quarks or (anti-)diquarks to hop in and out of $V$, or to be created and destroyed in hadronic clusters across the interface between $V$ and its complement $\overline{V}$. 
Simply tracing the density operator over the sector with $q_V = e \! \mod 3$,  without constraining the dynamics as well, is not sufficient because $q_V = e\! \mod 3$ must be maintained at all Euclidean times in the path-integral representation of the corresponding partition function. Intuitively, we have achieved this by inserting a static selective membrane between $V$ and $\overline{V}$, permeable for hadrons but not quarks, by prohibiting all processes that could change the center-electric flux through $S$. As a result, only the net baryon number is allowed to fluctuate in  $V$, but not the net quark number modulo three. 

The realization of such a membrane is conceptually rather simple: one first introduces closed center-vortex sheets between \emph{all} Euclidean time slices in the plaquette action such that the spatial links of the twisted plaquettes intersect  with $S=\partial V$, i.e.~belong to the coboundary  $S^\ast$ dual in three dimensions to $S$.
With $L_4$ time slices and $N_c$ colors there are in total $N_c L_4 $ independent combinations of such twists classified by sets $\{z_\tau\in Z_{N_c}\}$, $\tau = 0,\dots L_4-1$, of $N_c^\mathrm{th}$ roots of unity. The final path-integral representation of the partition functions $Z(q_V =_{N_c} e)$ with fixed net quark numbers $q_V =  e\! \mod N_c$ in $V$ is then obtained from $L_4$ discrete Fourier transforms over $\{z_\tau\}$, all with the same center-electric flux $e \in \mathbb Z_{N_c}$. 

The construction generalizes that of 't~Hooft's electric fluxes in the pure gauge theory, as $ Z_{N_c}$-Fourier transforms of ensembles with twisted boundary conditions. The boundary here is $S$, and the independent combinations $\{z_\tau\}$ of twists only modify those temporal plaquettes in the gauge action whose spatial links are in $S^*$. In particular, the fermion determinat remains unchanged. We have demonstrated this perhaps surprisingly simple result from two different approaches: a dualization of the fermionic action, where it arises quite naturally, and a transfer-matrix construction,  where we have to carefully construct a suitable transfer operator $\hat{T}$, to see this explicitly. The fermions are nevertheless essential in the construction because without them, in the pure gauge theory, the closed center-vortex sheets could all be removed from the action at no cost (which is why one uses 't~Hooft's twisted boundary conditions there instead).

Another important difference is that in the pure gauge theory the center-flux sectors are invariant under the dynamics \cite{BorgsSeiler1983}. 
Therefore, one only has to restrict the trace over the density operator to the corresponding flux sector. Without spatial hops in the static fermion determinant, this remains true in the heavy-dense limit of QCD
for the sectors with non-vanishing center charge in a subvolume $V$. In full QCD, on the other hand, in addition to restricting the trace to sectors with fixed net quark numbers $q_V = e\!\mod 3$, we have to modify the dynamics at the surface $S=\partial V$ as well, in order to prevent individual quarks from fluctuating in and out of $V$ as explained above. This is achieved by introducing the independent twists between \emph{all} Euclidean time slices.

Our work offers several interesting opportunities for further research: Explicitly being based on center charges and fluxes with their local $\mathbb Z_3$-Gauss law the construction can be used to study the percolation properties of the deconfinement transition \cite{Satz1998}.
Starting from our flux-tube model for heavy-dense QCD, 
one could revisit the idea going back to Patel \cite{Patel1984} that deconfinement can be described in terms of percolating clusters of flux strings. As we have seen in Sec.~\ref{sec::electrix_flux_ensembles}, the center-electric flux required for $q_V \not= 0\!\mod 3$ in $V$ at low temperatures only allows clusters of center charges that are strictly localized at the surface $S=\partial V$. With increasing temperature, however, the clusters extend further into the bulk and eventually, these localized structures completely dissolve. This transition is accompanied by percolating electric fluxes. For sufficiently heavy quarks, the percolation transition coincides with the first-order transition of the equivalent dual $Z_3$-Potts model. It is quite possible, however, that the percolation transition of the center-electric fluxes persists, as a second-order geometric phase transition without thermodynamic singularity, into the crossover region of the $Z_3$ symmetry at lower quark masses along a so-called  Kert\'esz line \cite{Kertesz1989,FortunatoSatz2001,FortunatoSatz2002}. At vanishing quark mass, the flux-tube model would then predict bond percolation. This is of course not relevant for QCD, but the Kert\'esz line in the plane of temperature and light-quark mass $m_q$ could in principle be studied by measuring the percolation  strength of the center-electric fluxes in full QCD on the lattice. In this way, one could address the question whether it might connect the deconfinement and chiral transitions for $m_q\to\infty $ and $m_q\to 0 $.  Perhaps it will not, but rather end for $m_q\to 0 $ in the somewhat higher transition temperature of a ``stringy liquid'' phase \cite{GlozmanPhilipsenPisarski2022} instead, where color-electric flux tubes are predicted to get screened.

Last but not least, understanding the sectors with different net quark numbers  $q_V = e\!\mod 3$ in the volume $V$ should allow to derive analogous path-integral representations for the reduced density matrices where the degrees of freedom in the complement $\overline V$ have been traced out while keeping the required flux through $S=\partial V$ fixed. This could then in turn be used to calculate the entanglement entropy between $V$ and $\overline V$ expected for the non-vanishing electric flux, and hence generalize existing studies \cite{AokiIizuka2017, BuividovichPolikarpov2008,BuividovichPolikarpov2008_2}  to the charged sectors in full QCD.

\newpage

\appendix
\section{Strong-coupling partition functions}
\label{sec::site-factors}

Naive center projection of the untraced straight Polyakov line $W_i \in \mathrm{SU}(3)$ at site $i$ would lead to
\begin{equation*}
 L_i = \tr W_i  \to 3 \, z_i \, ,\;\; z_i \in \text Z_3 = \left\{1,\, \e^{2 \pi \i/3},\, \e^{4 \pi \i/3} \right\}\, . %\label{eq::centerproj}
\end{equation*}
The static fermion determinant, with this center projection in (\ref{eq::statDet}), then reduces to factors of
\be
 Q(3z) = \big(1 + h z\big)^6 \big(1+\bar h z^*\big)^6 , \label{eq::statDetcp}
\ee
where the exponent of the quark and anti-quark factors corresponds to $2N_c N_f$ and simply counts the number of fermionic degrees of freedom per spatial site: $2$ for the two spin components, $N_c=3$ for the three colors of QCD, and $N_f = 1$ for one flavor, here. This already shows that center projection simply amounts to treating color as yet another quantum number, like flavor or spin, and in particular, without restriction to color singlets. While this might be acceptable for an effective Polyakov-loop theory at weak coupling, it disagrees with the strong-coupling limit, where the $\mathrm{SU}(3)$ group integrations of the Polyakov loops at the $N_s$ spatial sites in (\ref{eq::effact}) for $\lambda \to 0$, and $\bar h \to 0$ for $m \gg T$ with $\mu>0$, yield,
\begin{equation*}
Z_\mathrm{eff} \to (1+4 h^3 + h^6)^{N_s} %(1+4 \bar h^3 + \bar h^6)^{N_s}
\,. \label{eq::str-couplim_A}
\end{equation*}
The color-singlet three-quark states in the one-flavor theory are spin-$3/2$ baryons with a multiplicity of $4$, as counted by the degeneracy factor of the $h^3$ term. The correspondig center-projected site factor (\ref{eq::statDetcp}) would yield $(1 + 20h^3 + h^6)$, on the other hand. Although quark numbers would still be restricted to multiples of three, due to $\text Z_3$ symmetry, it would include additional 16 spin-1/2 color-octet three-quark states.

Modeling confinement as commonly done in Polyakov-loop enhanced quark models, via simply replacing the Polyakov-loop variables by their value at the center-symmetric maximum of the reduced Haar measure in the center of the fundamental domain is equally unacceptable here: With $L_i\to 0$ this would yield a fermionic site factor $(1 + 2h^3 + h^6)$, with a multiplicity of two for spin $1/2$ of the three-quark state in the one-flavor theory.
It might therefore be worthwhile to consider effective Polyakov-loop potentials that parametrize a transition between the trivial center element, with $L = 3 $,
and the $\text Z_3$ set of midpoints along the grey contour, as indicated by the double-headed arrow in Figure \ref{fig::redHaar}, when changing between weak and strong coupling or high and low temperatures. 
In particular, this could be used to define Roberge-Weiss symmetric Polyakov-loop enhanced quark models consistent with QCD at strong-coupling, instead of the usual ones which parametrize a transition between $L=3$ and the maximum of the reduced Haar measure at $L=0$ in the center of the fundamental domain.

%%%%%%%

In the strong-couling limit, $\lambda\sim\beta \to 0$ or $\sigma \to\infty$, the partition function, just
as the static fermion determinant, reduces to independent site-factors $S(h, \bar h)$,
\begin{equation*}
Z_\mathrm{eff} \to S(h,\bar h)^{N_s} \,.
\end{equation*}
With the full static determinant of (\ref{eq::statDet}) and $\mathrm{SU}(3)$ group integrations, we need moments $T_{m,n}$ of the reduced Haar measure up to $m+n=4$. The result is straightforwardly computed as
\begin{align}
S_{\mathrm{SU}(3)}(h,\bar h) &= 1+ 4 h\bar h + 4 h^3+ 4 \bar h^3  + 10 h^2 \bar h^2 \label{eq::SSU3} \\
&\hskip -.4cm + 6 h^4 \bar h + 6 h \bar h^4 + h^6 + 20 h^3 \bar h^3  + \bar h^6 \nonumber\\
&\hskip -.4cm + 6 h^5 \bar h^2 + 6 h^2 \bar h^5 + 10 h^4 \bar h^4 + 4 h^6 \bar h^3 + 4 h^3 \bar h^6 \nonumber\\
&\hskip -.4cm +  4 h^5 \bar h^5   + h^6 \bar h^6 \, .\nonumber
\end{align}
With our midpoint definition (\ref{eq::midpdef}) for the group integrations we obtain from our effective $\text Z_3$-Potts model (\ref{eq::effPotts}), or the corresoponding flux-tube model (\ref{eq::effFTM}), 
\begin{align}
S_\mathrm{Potts}(h,\bar h) &= 1+ 4 h\bar h + 4 h^3+ 4 \bar h^3  + 9 h^2 \bar h^2 \label{eq::SPotts} \\
&\hskip -.4cm + 6 h^4 \bar h + 6 h \bar h^4 + h^6 + 16 h^3 \bar h^3  + \bar h^6 \nonumber\\
&\hskip -.4cm + 6 h^5 \bar h^2 + 6 h^2 \bar h^5 + 9 h^4 \bar h^4 + 4 h^6 \bar h^3 + 4 h^3 \bar h^6 \nonumber\\
&\hskip -.4cm +  4 h^5 \bar h^5   + h^6 \bar h^6 \,  .\nonumber
\end{align}
The difference arises from $T_{2,2} = 2$, corresponding to the $2$ singlets in the $3\otimes 3\otimes \bar 3 \otimes \bar 3$ product representation of which we count only one,
\begin{align*}
  S_{\mathrm{SU}(3)}(h,\bar h) - S_\mathrm{Potts}(h,\bar h) &= h^2\bar h^2 + 4 h^3 \bar h ^3 + h^4\bar h^4 \\
  &\hskip -2cm = \e^{-4 m/T} + 4 \e^{-6 m/T} +  \e^{-8 m/T} \, . 
\end{align*}
This difference is purely mesonic and independent of $\mu $. Moreover, it vanishes when either $\bar h \to 0$ or $ h\to 0$, and hence does not influence the heavy-dense limit where $m\gg T$.  

The original flux-tube model of Ref.~\cite{CondellaDetar2000}, with two spin-degrees of freedom, corresponds to fermionic site factors 
of the form (with a power of 2 for the two spin sates):      
\begin{equation*}
\big( 1 + h^3 + \bar h^3 + (h + \bar h^2) z_i + (h^2+ \bar h) z_i^* \big)^2\, .  %\label{eq::sf-ftmorig}
\end{equation*}
Upon $\text Z_3$ summation, this yields
\begin{align*}
S_\mathrm{FT}(h,\bar h) &= 1+ 2 h\bar h + 4 h^3 + 4 \bar h^3  + 2  h^2 \bar h^2  \\
&\hskip .4cm + h^6 + 2 h^3 \bar h^3  + \bar h^6 \,  ,
\end{align*}
This reproduces the correct limits $\bar h \to 0$ or $ h\to 0$, but it is wrong already at the order $m+n = 2$, producing two mesonic quark-anti-quark states 
%with a degeneracy of 2 
for a single flavor 
where there should be one spin-0 and one spin-1 meson in the strong-coupling limit with the static fermion determinant, see Eqs.~(\ref{eq::SSU3}) and (\ref{eq::SPotts}).

Finally, the non-confining center projection from (\ref{eq::statDetcp}) with $\text Z_3$ summation gives,
\begin{align*}
S_\mathrm{CP}(h,\bar h) &= 1+ 36 h\bar h + 20 h^3 + 20 \bar h^3 + 225  h^2 \bar h^2  \\
&\hskip -1cm + 90 h^4 \bar h + 90 h \bar h^4  + h^6 + 400 h^3 \bar h^3 + \bar h^6  \\
&\hskip -1cm + 90 h^5\bar h^2 + 90 h^2 \bar h^5 + 225 h^4 \bar h^4  + 20 h^6 \bar h^3 + 20 h^3 \bar h^6  \\
&\hskip -1cm + 36 h^5\bar h^5 + h^6 \bar h^6
\,  .
\end{align*}

\section{External Fields}
\label{sec::appendix_external_fields}

In this section, we state the mapping from the flux tube parameters $\beta$, $\mu$ and $m$ to $\chi$ and the external fields $\eta$ and $\eta'$. The mapping is done completely analogously to the mapping described in \cite{CondellaDetar2000}. We only state the result. The on-site fermionic part is written as
\begin{equation*}
    Q(z) = a + bz + cz^\ast = \e^{\chi + \eta\mathrm{Re}z + \mathrm{i}\eta'\mathrm{Im}z}\,.
\end{equation*}
This implies
\begin{align*}
    \begin{split}
	\chi =&\frac{1}{3}\ln\left[a+b+c\right]\\
	+&\frac{1}{3}\ln\left[\left(a - \frac{1}{2}(b+c)\right)^2+\frac{3}{4}(b-c)^2\right]\,,
	\end{split}\\
	\begin{split}
	\eta = &\frac{2}{3}\ln\left[a+b+c\right]\\
	-&\frac{1}{3}\ln\left[\left(a - \frac{1}{2}(b+c)\right)^2+\frac{3}{4}(b-c)^2\right]\,,
	\end{split}\\
	\eta' = &\frac{2}{\sqrt{3}}\arctan\left(\frac{\sqrt{3}(b-c)}{2a-(b+c)}\right)
\end{align*}
with
\begin{align*}
	\begin{split}
		a = &1+4\overline{h}^3+\overline{h}^6+4\overline{h}h+6\overline{h}^4h+9\overline{h}^2h^2\\
		+&6\overline{h}^5h^2+4h^3+16\overline{h}^3h^3+4\overline{h}^6h^3+6\overline{h}h^4\\
		+&9\overline{h}^4h^4+6\overline{h}^2h^5+4\overline{h}^5h^5+h^6+4\overline{h}^3h^6+\overline{h}^6h^6\,,
	\end{split}\\
	\begin{split}
		b = &3\overline{h}^2+2\overline{h}^5+2h+8\overline{h}^3h+2\overline{h}^6h+6\overline{h}h^2\\
 		+&9\overline{h}^4h^2+12\overline{h}^2h^3+8\overline{h}^5h^3+3h^4+12\overline{h}^3h^4\\
		+&3\overline{h}^6h^4+4\overline{h}h^5+6\overline{h}^4h^5+3\overline{h}^2h^6+2\overline{h}^5h^6\,,
	\end{split}
\end{align*}
and
\begin{align*}
	\begin{split}
    	c = &2\overline{h}+3\overline{h}^4+6\overline{h}^2h+4\overline{h}^5h+3h^2+12\overline{h}^3h^2\\
    	+&3\overline{h}^6h^2+8\overline{h}h^3+12\overline{h}^4h^3+9\overline{h}^2h^4+6\overline{h}^5h^4\\
    	+&2h^5+8\overline{h}^3h^5+2\overline{h}^6h^5+2\overline{h}h^6+3\overline{h}^4h^6\,.
	\end{split}
\end{align*}

\section{Transfer Matrix Construction}
\label{sec::appendix_trans_const}

Complementary to Section \ref{sec::trans_const} we here give a detailed construction of the transfer matrix operator and a proof of its equivalence to the LQCD partition function. The transfer matrix operator $\hat{T}$ is constructed similar to \cite{Luescher1977} but with the difference of being asymmetric. This choice is important because it allows to restrict the influence of the projection operator $\hat{P}_{\phi_S}(-e)$ to the gauge action only.

The operator $\hat{T}$ is constructed in a similar manner as in \cite{Luescher1977} but the construction was modified such that the corresponding integral kernel $K(U,U')$ is asymmetric, i.e. the integral kernel has the form
\begin{equation*}
    K(U,U') = S(U,U')T_G(U')T_F(U')
\end{equation*}
with the spatial plaquette part
\begin{align*}
    T_G(U') &=\exp\left(\frac{2}{g_0^2}\sum_{i,k<m}\ReTr\left(U'_{\partial p_{i,km}}\right)\right)\,,\\
\end{align*}
the timelike plaquettes part
\begin{align*}
	S(U,U')&=\exp\left(\frac{2}{g_0^2}\sum_{i,k}\ReTr\left(U_{\langle i,i+\hat{k}\rangle}U'_{\langle i+\hat{k},i\rangle}\right)\right)
	\, ,
\end{align*}
and the fermionic part 
\begin{widetext}
\begin{equation*}
    T_F(U') = (2\kappa)^{12 L_1L_2L_3}\det\left(B(U')\right)\exp\left(\hat{\xi}^\dagger_\mathrm{q} c(U')\hat{\xi}^\dagger_{\overline{\mathrm{q}}}\right)\exp\left(\hat{\xi}^\dagger_\mathrm{q} \ln\left(B(U')\right)\hat{\xi}_\mathrm{q} -\hat{\xi}_{\overline{\mathrm{q}}} \ln\left(B(U')\right)\hat{\xi}^\dagger_{\overline{\mathrm{q}}} \right)\exp\left(\hat{\xi}_{\overline{\mathrm{q}}}c(U')\hat{\xi}_{\mathrm{q}}\right)\,.
\end{equation*}
The factors $T_G(U')$ and $S(U,U')$ will give the Wilson plaquette action later on and $T_F(U')$ will yield the fermion determinant. The matrices $c$ and $B$ are defined as rescaled versions of the matrices in \cite{Luescher1977}:
\begin{align}
    \label{eq::matrix_b}
    B_{ia\gamma,jb\gamma'} &= \frac{1}{2\kappa}\delta_{ij}\delta_{ab}\delta_{\gamma\gamma'}-\frac{1}{2}\delta_{\gamma\gamma'}\sum_{k}\left[(U_{\langle i, i+\hat{k}\rangle})^{ab}\delta_{i+\hat{k},j}+(U_{\langle j+\hat{k},j\rangle})^{ab}\delta_{j+\hat{k},i}\right]\,,\\
	c_{ia\gamma,jb\gamma'} &= \frac{1}{2}\sum_k \left[(U_{\langle i,i+\hat{k}\rangle})^{ab}(\mathrm{i}\sigma_k)_{\gamma\gamma'}\delta_{i+\hat{k},j}-(U_{\langle j+\hat{k},j\rangle})^{ab}(\mathrm{i}\sigma_k)_{\gamma\gamma'}\delta_{j+\hat{k},i}\right]\,,
\end{align}
where $\sigma_k$ are the Pauli spin matrices and $\gamma,\gamma' \in\{\uparrow,\downarrow\}$. By following the line of argumentation in \cite{Luescher1977,Palumbo2002,Mitrjushkin2002,Mitrjushkin2003}, the path integral formulation of Equation (\ref{eq::partition_trans_const}) can be computed explicitly. We write the trace as an integral over products of the kernels $K(U,U')$ and insert fermionic coherent states:
\begin{equation*}
    Z = \int\mathcal{D}U\mathcal{D}\Omega\mathcal{D}\overline{\xi}\mathcal{D}\xi\,\e^{-\overline{\xi}\xi}\bra{\e^{\beta\mu}\xi^{(0)}_{\mathrm{q}},\e^{-\beta\mu}\xi^{(0)}_{\overline{\mathrm{q}}}}K(U_0,U_1,\Omega)\ket{\Omega\xi^{(1)}_{\mathrm{q}},\xi^{(1)}_{\overline{\mathrm{q}}}\Omega^\dagger}\prod_{\tau = 1}^{L_4-1}\bra{\xi^{(\tau)}}K(U_\tau, U_{\tau+1})\ket{\xi^{(\tau+1)}}
\end{equation*}
with the integral kernel $K(U_0,U_1,\Omega)$ where all $U_0$-dependencies and the creation and annihilation operators are transformed by the gauge transformation $\hat{\varrho}(\Omega)$. The computation of the expectation values are straightforward and result in
\begin{equation*}
    Z \propto \int\mathcal{D}U\mathcal{D}\Omega\mathcal{D}\overline{\xi}\mathcal{D}\xi\,\e^{S_G(U,\Omega)}\e^{S_F(U,\Omega,\overline{\xi},\xi)}\,.
\end{equation*}
At the end, we will identify $\Omega_i$ as gauge fields in direction of Euclidean time. Thus the pure gauge part $S_G(U,\Omega)$ will become the Wilson plaquette action. The fermionic part is given by
\begin{equation*}
\begin{split}
    S_F(U,\Omega,&\overline{\xi},\xi) = -\sum_{\tau = 0}^{L_4-1}\left(\overline{\xi}^{(\tau)}_\mathrm{q}\left[\delta_{\tau,0}\e^{-\beta\mu}\Omega^\dagger + (1-\delta_{\tau,0})\right]\xi^{(\tau)}_\mathrm{q} - \xi_{\overline{\mathrm{q}}}^{(\tau)}\left[\delta_{\tau,0}\e^{\beta\mu}\Omega + (1-\delta_{\tau,0})\right]\overline{\xi}_{\overline{\mathrm{q}}}^{(\tau)}\right)\\
    +&\sum_{\tau=0}^{L_4-1}\overline{\xi}_\mathrm{q}^{(\tau)}c(U_{\tau+1})\overline{\xi}_{\overline{\mathrm{q}}}^{(\tau)}+\sum_{i=0}^{L_4-1}\xi_{\overline{q}}^{(\tau)}c(U_\tau)\xi_{\mathrm{q}}^{(\tau)}+\sum_{\tau=0}^{L_4-1}\bigg(\overline{\xi}^{(\tau)}_\mathrm{q} B(U_{\tau+1})\xi^{(\tau+1)}_\mathrm{q}-\xi^{(\tau+1)}_{\overline{\mathrm{q}}}B(U_{\tau+1})\overline{\xi}^{(\tau)}_{\overline{\mathrm{q}}}\bigg)\,.
\end{split}
\end{equation*}
\end{widetext}
This action can be transformed into the Hasenfratz-Karsch action. For this purpose we shift the Grassmann variables in direction of Eucldiean time:
\begin{equation*}
    \begin{split}
        \overline{\xi}^{(\tau)}_{\mathrm{q}}&\longrightarrow\overline{\xi}^{(\tau+1)}_{\mathrm{q}}\,,\quad\overline{\xi}^{(L_4-1)}_{\mathrm{q}}\longrightarrow -\overline{\xi}^{(0)}_{\mathrm{q}}\,,\\
        \overline{\xi}^{(\tau)}_{\overline{\mathrm{q}}}&\longrightarrow\overline{\xi}^{(\tau+1)}_{\overline{\mathrm{q}}}\,,\quad
        \overline{\xi}^{(L_4-1)}_{\overline{\mathrm{q}}}\longrightarrow -\overline{\xi}^{(0)}_{\overline{\mathrm{q}}}
    \end{split}
\end{equation*}
for $\tau < L_4-1$. The terms of the first sum of $S_F$ become the hopping terms in direction of Euclidean time and the terms of the last sum become part of the spatial hops. 

Afterwards, the Dirac spinors are introduced:
\begin{align*}
	(\overline{\xi}_{\mathrm{q}}^{\uparrow})_{i,a}^{(\tau)} \longrightarrow \overline{\psi}_{(i,\tau)}^{a1}\,, &\qquad (\xi^{\uparrow}_{\mathrm{q}})_{i,a}^{(\tau)}\longrightarrow \psi_{(i,\tau)}^{a1}\,,\\
	(\overline{\xi}_{\mathrm{q}}^{\downarrow})_{i,a}^{(\tau)} \longrightarrow \overline{\psi}_{(i,\tau)}^{a2}\,, &\qquad (\xi^{\downarrow}_{\mathrm{q}})_{i,a}^{(\tau)}\longrightarrow \psi_{(i,\tau)}^{a2}\,,\\
	(\overline{\xi}_{\overline{\mathrm{q}}}^{\uparrow})_{i,a}^{(\tau)} \longrightarrow \psi_{(i,\tau)}^{a3}\,, &\qquad (\xi^{\uparrow}_{\overline{\mathrm{q}}})_{i,a}^{(\tau)}\longrightarrow \overline{\psi}_{(i,\tau)}^{a3}\,,\\
	(\overline{\xi}_{\overline{\mathrm{q}}}^{\downarrow})_{i,a}^{(\tau)} \longrightarrow \psi_{(i,\tau)}^{a4}\,, &\qquad (\xi^{\downarrow}_{\overline{\mathrm{q}}})_{i,a}^{(\tau)}\longrightarrow \overline{\psi}_{(i,\tau)}^{a4}\,.
\end{align*}
To finally gain the Hasenfratz-Karsch action in maximal temporal gauge, the fields are scaled suitably and we identify $\Omega_{i} = U_{\langle (i,0),(i,1)\rangle}$. Consequently the asymmetric transfer operator yields the LQCD partition function. 

\section{Euclidean Time Continuum}
\label{sec::euclidean_time_continuum}

The asymmetric transfer operator $\hat{T}$ is shown to be given by a gauge-invariant Hamilton operator $\hat{H} = \hat{H}_{\mathrm{G}} + \hat{H}_{\mathrm{F}}$ in linear order of Euclidean spacing $a_4$, i.e. 
\begin{equation*}
    \hat{T}_{a_4} = \mathds{1} - a_4\hat{H} +\mathcal{O}(a_4^2)\,.
\end{equation*}
To consider the limit, we have to introduce the Euclidean spacing $a_4$ into the transfer operator properly. The lattice spacing is set to $a$ in spatial directions. The pure gauge action and fermion action are given by (\cite{Smit2002})
\begin{align*}
    \begin{split}
    S_G &= \frac{a}{a_4}\frac{2}{g_0^2}\sum_{x, k}\ReTr U_{\partial p_{x,k4}}\\
    &\hphantom{\frac{a}{a_4}\frac{2}{g_0^2}\sum_{x, k}}+\frac{a_4}{a}\frac{2}{g_0^2}\sum_{x, k < m}\ReTr U_{\partial p_{x,km}}\,,
    \end{split}\\
    \begin{split}
        S_F &= s(a_4,a)\sum_{x}\bigg[\overline{\psi}_x\Lambda_{x,4}^+\psi_{x+\hat{4}}+ \overline{\psi}_{x+\hat{4}}\Lambda_{x,4}^-\psi_x\bigg]\\
        &+\frac{a_4}{a}s(a_4,a)\sum_{x,k}\bigg[\overline{\psi}_x\Lambda_{x,k}^+\psi_{x+\hat{k}}+ \overline{\psi}_{x+\hat{k}}\Lambda_{x,k}^-\psi_x\bigg]\\
        &-\sum_x\overline{\psi}_x\psi_x
    \end{split}
\end{align*}
with
\begin{equation*}
    s(a_4,a) = \frac{1}{2a_4(m+3a^{-1}) + 2}\,.
\end{equation*}
For equal spacing $a_4 = a$, we get Equation (\ref{eq::plaquette_action}) and (\ref{eq::fermion_action}). We introduce the lattice spacing in the transfer operator $\hat{T}_{a_4}$. We insert the factor $\frac{a_4}{a}$ into the exponent of $T_G(U')$ and the factor $\frac{a}{a_4}$ into the exponent of $S(U,U')$. The fermionic part is modified in the following way:
\begin{align*}
    (2\kappa)^{12L_1L_2L_3} &\longrightarrow (2s(a_4,a))^{12L_1L_2L_3}\,,\\
    c(U') &\longrightarrow \frac{a_4}{a}c(U')\,,\\
    B(U') &\longrightarrow \frac{a_4}{a}B(U')
\end{align*}
with the matrix $B$ modified such that
\begin{equation*}
    B = \frac{a}{2a_4s(a_4,a)}\mathds{1} + R\,,
\end{equation*}
where $R$ is the non-diagonal part of Equation (\ref{eq::matrix_b}). This transfer operator gives the path integral with the above action as can be verified by following the lines of argument in Appendix \ref{sec::appendix_trans_const}. Furthermore, we have $\hat{T}_{a_4 = a} = \hat{T}$. Hence, $\hat{T}_{a_4}$ gives the correct $a_4$-dependency. The transfer operator can be decomposed into the product of three operators:
\begin{equation*}
    \hat{T}_{a_4} = \hat{S}_{a_4}\hat{T}_{G,a_4}\hat{T}_{F,a_4}\,.
\end{equation*}
The operator $\hat{S}_{a_4}$ is the convolution operator of time-like plaquettes  and $\hat{T}_{G,a_4}$ is the operator multiplying a state with $T_G(U)$. These two operators only act on the gauge part of a state. For small $a_4$, their approximation is derived in Ref. \cite{BorgsSeiler1983} and we have
\begin{equation*}
    \hat{S}_{a_4}\hat{T}_{G,a_4} = \mathds{1} - a_4 \hat{H}_{\mathrm{G}} + \mathcal{O}(a_4^2)\,.
\end{equation*}
The operator $\hat{T}_{F,a_4}$ multiplies a state by $T_F(U)$. The expansion in orders of $a_4$ is done by expanding all factors separately which is a straightforward task. We end with the Hamiltonian 
\begin{equation*}
    \begin{split}
        \hat{H}_{\mathrm{F}} = -&\hat{\xi}_{\mathrm{q}}^\dagger\left(m+\frac{3}{a}\right)\hat{\xi}_{\mathrm{q}}
        +\hat{\xi}_{\overline{\mathrm{q}}}\left(m+\frac{3}{a}\right)\hat{\xi}_{\overline{\mathrm{q}}}^\dagger \\
        +&\frac{1}{a}\left(\hat{\xi}_{\overline{\mathrm{q}}}R(U)\hat{\xi}_{\overline{\mathrm{q}}}^\dagger-\hat{\xi}_{\mathrm{q}}^\dagger R(U)\hat{\xi}_{\mathrm{q}}\right)\\
        -&\frac{1}{a}\left(\hat{\xi}^\dagger_{\mathrm{q}}c(U)\hat{\xi}_{\overline{\mathrm{q}}}^\dagger+\hat{\xi}_{\overline{\mathrm{q}}}c(U)\hat{\xi}_{\mathrm{q}}\right)\,.
    \end{split}
\end{equation*}
Finally we introduce the creation and annihilation operators
\begin{align*}
	(\hat{\chi}_{i})_{1,a} &= (\hat{\xi}_{\mathrm{q}}^{\uparrow})_{i,a}\,,   &  (\hat{\chi}_{i})_{3,a} &= (\hat{\xi}_{\overline{\mathrm{q}}}^{\uparrow})^\dagger_{i,a}\,,\\
	(\hat{\chi}_{i})_{2,a} &= (\hat{\xi}_{\mathrm{q}}^{\downarrow})_{i,a}\,, &  (\hat{\chi}_{i})_{4,a} &= (\hat{\xi}_{\overline{\mathrm{q}}}^{\downarrow})_{i,a}^\dagger\,.
\end{align*}
This gives
\begin{equation*}
    \begin{split}
    &\hat{H}_{\mathrm{F}} =
    \frac{1}{a}\sum_{i,k}\bigg[\hat{\chi}_{i}^\dagger\gamma_4 P_- U_{\langle i, i+\hat{k}\rangle}\hat{\chi}_{i+\hat{k}} \\ &\,+\hat{\chi}^\dagger_{i+\hat{k}}\gamma_4P_+ U_{\langle i+\hat{k}, i\rangle}\hat{\chi}_{i}\bigg]
    -\hat{\chi}^\dagger\left[\gamma_4\left(m+\frac{3}{a}\right)\right]\hat{\chi}
    \end{split}
\end{equation*}
with $P_{\pm} = \frac{1}{2}(1\pm\gamma_4)$. This is just the Hamilton operator of Wilson fermions on the lattice \cite{MontvayMuenster1994}.

Finally, we can also argue that the modified transfer operator $\hat{T}'$ has also a well-defined Euclidean time continuum with an hermitian, gauge-invariant Hamiltonian $\hat{H}'$. Therefore, the quantum mechanical interpretation also holds for the modified system. The modified transfer operator $\hat{T}'_{a_4}$ with explicit $a_4$-dependency gives
\begin{align*}
    (\hat{T}'_{a_4})^{L_4} &= \left(\frac{1}{3}\sum_{z\in\mathrm{Z}_3}\hat{\phi}_S^z \hat{T}_{a_4}\hat{\phi}_S^{z^{-1}}\right)^{L_4}\\
    &= \left(\mathds{1} -a_4\left[\frac{1}{3}\sum_{z\in\mathrm{Z}_3}\hat{\phi}_S^z \hat{H}\hat{\phi}_S^{z^{-1}}\right]+\mathcal{O}(a_4^2)\right)^{L_4}\,.
\end{align*}
Hence, we read off the modified Hamiltonian
\begin{equation*}
    \hat{H}' = \frac{1}{3}\sum_{z\in\mathrm{Z}_3}\hat{\phi}_S^z \hat{H}\hat{\phi}_S^{z^{-1}}\,.
\end{equation*}
The operator $\hat{H}'$ is obviously hermitian and gauge-invariant because $\hat{H}$ is. 
\bibliography{references}

\end{document}